\documentclass[12pt]{iopart}
\usepackage{iopams} 
\usepackage{setstack}
\usepackage{graphicx}
\usepackage{caption}
\usepackage{subcaption}
\usepackage[colorlinks=true,linkcolor=blue,citecolor=blue,urlcolor=blue]{hyperref}

\usepackage{microtype}

\usepackage{braket}

\begin{document}

\title{Generating and protecting correlated quantum states under collective dephasing}

\author{Edoardo G. Carnio$^{1,2}$, Andreas Buchleitner$^{1,3}$, and Manuel Gessner$^{1,4,5}$}
\address{$^1$ Physikalisches Institut, Albert-Ludwigs-Universit\"at Freiburg,
Hermann-Herder-Stra\ss e 3, 79104 Freiburg, Germany}
\address{$^2$ Department of Physics, University of Warwick, Coventry, CV4 7AL, United Kingdom}
\address{$^3$ Freiburg Institute for Advanced Studies,
Albert-Ludwigs-Universit\"at Freiburg, Albertstra\ss e 19, 79104 Freiburg,Germany}
\address{$^4$ QSTAR, CNR-INO, and LENS, Largo Enrico Fermi 2, I-50125 Firenze, Italy}
\address{$^5$ Istituto Nazionale di Ricerca Metrologica (INRiM), I-10135 Torino, Italy}

%
%

\begin{abstract}
We study the collective dephasing process of a system of non-interacting atomic qubits, immersed in a spatially uniform magnetic field of fluctuating intensity. The correlation properties of bipartite states are analysed based on a geometric representation of the state space. Particular emphasis is put on the dephasing-assisted generation of states with a high correlation rank, which can be related to discord-type correlations and allow for direct applications in quantum information theory. Finally we study the conditions that ensure the robustness of initial entanglement and discuss the phenomenon of time-invariant entanglement.
\end{abstract}

\section{Introduction}
Ensembles of trapped, laser-cooled atomic particles provide some of the best-controlled experimental platforms to study quantum dynamics, to engineer effective interactions, or to generate specific quantum states \cite{wineland98,morsch,ciraczoller,hartmut,gross10,ben,monz13,gross15}. An essential requirement for such levels of control is the efficient isolation of the system from its environment, whose detrimental influence leads to the decay of coherent superpositions \cite{breuerbook}. The loss of coherence often also implies the loss of quantum correlations, such as entanglement, which are required, e.g., to process quantum information \cite{horodecki^4,nielsen}.

One persistently dominant source of error, common to most experiments on trapped atomic particles, is caused by intensity fluctuations of external electromagnetic fields, which are needed to lift degeneracies, to compensate background fields, or to manipulate the quantum state of the system \cite{gross10,monz13}. Since such fields are typically generated by large coils outside the vacuum chamber, the resulting field is spatially homogeneous along the positions of the trapped particles. The unavoidable fluctuations of the field strength therefore lead to a \emph{collective} dephasing process, which is formally described by an ensemble average over the fluctuating parameter \cite{ben,Benatti,GB13,matteos,letter,PhDGessner}. The collective nature of this noise provides new possibilities to protect coherent superpositions \cite{letter}, and, as demonstrated in a recent experiment \cite{ben}, to generate robust, strongly correlated, albeit separable quantum states.

In this article we study the impact of the collective dephasing process on different types of correlations inscribed into quantum states, using the analytical description of the collective dephasing dynamics outlined in \cite{letter}. Our analysis focuses on the correlation rank, which in turn entails direct consequences for the discord-type correlations \cite{Gessner:2012fk}. Specifically, in Sec.~\ref{sec:generate}, we study how strongly correlated two-qubit states can be generated via collective dephasing, as a function of tunable external control parameters, e.g., the magnetic field direction. We discuss specific applications of the produced quantum states in the context of quantum information protocols. In Sec.~\ref{sec:generalk} we follow a complementary approach and analyse the protection of existing correlations during the collective dephasing process. We further discuss robust conditions that lead to the striking phenomenon of \textit{time-invariant} entanglement: the perfect conservation of the initial entanglement, even for states whose purity is reduced due to the dephasing process.

\section{Correlations in quantum states}\label{sec:correlations}
\subsection{Entanglement, discord and correlation rank}\label{sec.correlationsintro}
We begin by reviewing different notions of correlations in quantum states that will become relevant in the course of this article. \textit{Quantum entanglement} captures the non-classical correlations of a quantum state in a composite Hilbert space $\mathcal{H}^{A}\otimes\mathcal{H}^{B}$, i.e., it expresses the inability to characterize the full quantum state $\rho$ via a classical probability distribution $p_i$ and local density operators $\{\rho^{A}_i\}_i$ and $\{\rho^{B}_i\}_i$ on the local Hilbert spaces $\mathcal{H}^{A}$ and $\mathcal{H}^{B}$, respectively. In particular, a quantum state is \textit{separable} (i.e., not entangled) if it can be expressed as a convex linear combination of product states \cite{horodecki^4,werner,Mintert},
\begin{eqnarray}
\rho_s = \sum_i p_i \rho^{A}_i \otimes \rho^{B}_i.
\end{eqnarray}

Determining whether a given mixed quantum state is entangled, or evaluating an appropriate measure to quantify entanglement, is generally a very hard task \cite{Mintert,horodecki^4}. Only for the special case of two-qubit systems ($\mathcal{H}^A=\mathcal{H}^B=\mathbb{C}^2$), an algebraic expression that exactly quantifies the entanglement of arbitrary mixed states is available \cite{wootters}. This measure, the \textit{concurrence}, is determined on the basis of the eigenvalues $\lambda_i$ of $\sqrt{\sqrt{\rho} \tilde{\rho}\sqrt{\rho}}$, labelled in decreasing order, where the spin-flipped state $\tilde \rho = (\sigma_y \otimes \sigma_y) \rho^* (\sigma_y \otimes \sigma_y)$ is obtained by collective application of the Pauli matrix $\sigma_y$ to $\rho^*$, and complex-conjugation is performed in the computational basis. The concurrence of the state $\rho$ is then given by $C(\rho) = \max \lbrace 0, \lambda_1 - \lambda_2 -\lambda_3-\lambda_4 \rbrace$. For higher dimensional problems, only algebraic lower bounds are available \cite{Mintert}.

The \emph{quantum discord} describes the disturbance of local measurements on quantum states of composite systems \cite{Modi}. As we will see in the following, these features related to the quantum-mechanical measurement process can be expressed via non-vanishing commutators, and can only occur in correlated quantum states. They can, however, also manifest in separable states that exhibit only classical correlations, and therefore describe a more general type of quantum properties than entanglement. Formally, a state $\rho$ has zero discord if it can be written as \cite{Modi}
\begin{eqnarray}\label{eq.zerodiscord}
\rho_c = \sum_i p_i \ket{\psi_i^A}\bra{\psi_i^A} \otimes \rho^B_i,
\end{eqnarray}
where $\lbrace \psi_i^A \rbrace_i$ is an orthonormal basis of $\mathcal{H}^A$. This definition is equivalent to the following: a state $\rho$ has zero discord if and only if there exists a non-selective local projective measurement on $\mathcal{H}^A$ that leaves the state invariant, i.e., $\rho = \sum_i (\Pi_i^A\otimes \mathbb{I}^B)\rho(\Pi_i^A \otimes \mathbb{I}^B)$, where $\Pi_i^A = \ket{\psi_i^A}\bra{\psi_i^A}$. The definition presented here considers measurements on $\mathcal{H}^{A}$ and is straight-forwardly extended to measurements on $\mathcal{H}^{B}$. Due to the asymmetry of the definition, one should always specify in which subsystem measurements are performed, when discussing quantum discord.

Notice that every zero-discord state is separable but the converse is not true. The two concepts coincide only in the case of pure states. In contrast to entanglement, local operations on one of the subsystems can generate discord \cite{ben,PhysRevLett.105.190502}, which confirms that discord is not a proper measure for correlations. To quantify the correlations of a bipartite quantum state, we employ the rank of an appropriately constructed correlation matrix (the \emph{correlation rank}), which is the minimal number of bipartite operator products needed to describe the density operator \cite{PhDGessner,Gessner:2012fk}. For the formal definition, we write the density operator $\rho$ in an arbitrary basis of local Hermitian operators $\lbrace A_i\rbrace_i$ and $\lbrace B_j \rbrace_j$ \cite{PhysRevLett.105.190502}:
\begin{eqnarray}
\rho = \sum_{i=1}^{d_A^2} \sum_{j=1}^{d_B^2} r_{ij }A_i \otimes B_j,
\end{eqnarray}
where $d_{A,B} = \dim \mathcal{H}^{A,B}$. Here $R = (r_{ij})$ is the correlation matrix, a real-valued $d_A^2 \times d_B^2$-dimensional matrix whose rank $L$ is then called the correlation rank \cite{Gessner:2012fk}. Employing a singular value decomposition, we find non-zero singular values $\lbrace c_1,\ldots,c_L\rbrace$ and orthogonal matrices $U=(u_{ij})$ and $V=(v_{ij})$ such that $R = U \mathrm{diag}\left(c_1 , \ldots, c_L \right) V^T$. We obtain
\begin{eqnarray}\label{eq.svd}
\rho=\sum_{i,j} r_{ij}A_i \otimes B_j = \sum_{i,j} \sum_{k=1}^L u_{ik} c_k v_{jk} A_i \otimes B_j = \sum_{k=1}^L  c_k S_k \otimes F_k,
\end{eqnarray}
where $S_k = \sum_i u_{ik}A_i$ and $F_k = \sum_j v_{jk} B_j$. The above decomposition can be regarded as a Schmidt decomposition of a density matrix in terms of local operator bases \cite{PhDGessner}. A more familiar application of the Schmidt decomposition is known for pure bipartite quantum states, which are decomposed in terms of local vectors \cite{horodecki^4}. The associated singular value decomposition yields the Schmidt rank, which quantifies how entangled a pure state is \cite{horodecki^4}. Consequently, the correlation rank stands in close analogy to the Schmidt rank. In general, the correlation rank does not quantify entanglement but rather total correlations, i.e., any incompatibility with an uncorrelated product state, without an explicit distinction between classical and quantum nature of the correlations. In the special case of a pure state one obtains $L=\mathcal{S}^2$, where $\mathcal{S}$ denotes that state's Schmidt rank \cite{PhDGessner}.

Conclusions about the local quantum nature of the state can be drawn from the correlation rank by realizing that its maximal value is bounded from above for all zero-discord states. Generally, the correlation rank cannot exceed $d_\mathrm{min}^2$, where $d_{\mathrm{min}} = \min\lbrace d_A, d_B \rbrace$. However, as can be seen from the definition~(\ref{eq.zerodiscord}), states of zero discord with respect to measurements on $\mathcal{H}^{A,B}$ are represented in terms of local projectors $\ket{\psi_i^{A,B}}\bra{\psi_i^{A,B}}$ onto orthogonal subspaces. Since there exist no more than $d_{A,B}$ orthogonal subspaces in $\mathcal{H}^{A,B}$, the correlation rank of zero-discord states is bounded by $L \leq d_\mathrm{min}$. 

This is directly related to the observation that the operators $S_i$, which appear in Eq.~(\ref{eq.svd}), can be used to assess the discord of $\rho$: the state $\rho$ has zero discord (with respect to measurements on $\mathcal{H}^{A}$) if and only if all of the $S_i$ commute \cite{PhysRevLett.105.190502}. While local operations cannot increase the correlation rank $L$ \cite{Gessner:2012fk}, they can change the commutativity of the local operators $S_i$, and thereby generate discord \cite{PhysRevLett.105.190502} without actually generating correlations \cite{Gessner:2012fk}.

The correlation rank allows to distinguish separable states with high correlation rank ($d_\mathrm{min} < L \leq d_\mathrm{min}^2$) from those that can be generated from states of zero discord via local operations, as characterized by a low correlation rank ($L \leq d_\mathrm{min}$). Separable operations of the form $\sum_ip_i\Phi_i^A\otimes\Phi_i^B$, with local operations $\Phi^A$ and $\Phi^B$, can generate classical correlations among the particles and, consequently, are able to increase the correlation rank \cite{ben}. The collective dephasing operation to be discussed in this article represents such a separable operation.

\subsection{Representation of bipartite states}\label{sec:unitary-equivalence}
The density operator of any bipartite system can be represented in terms of the operator bases $\lbrace \mathbb{I}_{d_A}, \boldsymbol{\sigma}_A \rbrace \otimes \lbrace \mathbb{I}_{d_B}, \boldsymbol{\sigma}_B \rbrace$, where $\boldsymbol{\sigma}_{A,B}$ denote vectors whose entries are the generators of SU($d_{A,B}$), and $\mathbb{I}_{d_{A,B}}$ denote the respective identity operators. We obtain the state's \emph{Fano form} as \cite{fano,geometry}:
\begin{eqnarray}\label{eq:Fano-form}
\fl \rho = \frac{1}{d_A d_B} \left(\mathbb{I}_{d_Ad_B} + \bi r_A\cdot \boldsymbol{\sigma}_A \otimes \mathbb{I}_{d_B} + \mathbb{I}_{d_A} \otimes \bi r_B \cdot \boldsymbol{\sigma}_B + \sum_{i=1}^{d_A^2-1}\sum_{j=1}^{d_B^2-1} \beta_{ij} (\boldsymbol{\sigma}_A)_i \otimes (\boldsymbol{\sigma}_B)_j \right),
\end{eqnarray}
where $\bi r_A$ and $\bi r_B$ are the (generalized) Bloch vectors of the reduced subsystems, and $\beta$ is a real $(d_A^2-1)\times (d_B^2-1)$ matrix that describes the correlations between the subsystems. Because the state is completely characterized by $\bi r_A$, $\bi r_B$ and the $\beta$ matrix, throughout this article we will use the compact notation $\rho \doteq (\bi r_A, \bi r_B, \beta)$ \cite{horodecki96}. From the representation (\ref{eq:Fano-form}) it immediately follows that the correlation matrix of $\rho$ is given by
\begin{eqnarray}
R = \frac{1}{d_Ad_B} \pmatrix{
1 & \bi r_B^T \cr
\bi r_A & \beta
} ,
\end{eqnarray}
whose rank rk is \cite{meyer}
\begin{eqnarray}\label{eq:corr-rank}
L = \mathrm{rk} (R) = 1+\mathrm{rk}\left( \beta - \bi r_A \otimes \bi r_B \right).
\end{eqnarray}
While for a rigorous proof of the above identity we refer to Ref.~\cite{meyer}, we remark that the result can be obtained via block-diagonalization of the matrix $R=P\mathrm{diag}(1,M)Q$, where $P$ and $Q$ are rank-4 matrices and, in this case, $M=\beta - \bi r_A \otimes \bi r_B$ is a $3\times3$ matrix called the Schur complement (of the submatrix $1$).

In the remainder of this article, we mostly employ the reduced Bloch vectors and the $\beta$-matrix to investigate the impact of the collective dephasing process, which allows for an intuitive geometric description. A tool that we often employ to simplify our analysis are local \emph{unitary} transformations, since these affect neither the state's entanglement properties (by definition \cite{horodecki^4,Mintert}) nor its correlation rank (as we show explicitly later). Following \cite{horodecki96}, we  consider transformations $U = U_A \otimes U_B$ such that $\rho ' = U \rho U^\dagger$. For every unitary transformation $U_A$, there exists a rotation $O_A$ on the respective (generalized) Bloch sphere such that $U_A \left(\bi v \cdot \boldsymbol{\sigma}_A \right) U_A^\dagger = \left(O_A \bi v \right) \cdot \boldsymbol{\sigma}_A$. We then obtain the following transformation rules:
\begin{eqnarray}\label{eq.transform}
\bi r'_A = O_A \bi r_A,\quad \bi r'_B = O_B \bi r_B,\quad \beta' = O_A \beta O_B^T  .
\end{eqnarray}
If two matrices $\beta$ and $\beta'$ can be transformed into each other by such an operation, we write $\beta' \sim \beta$. In particular, we can always diagonalize the $\beta$ matrix by applying unitary transformations to the underlying quantum state.

We now show that such transformations do not change the rank of the correlation matrix. If $M = \beta - \bi r_A \otimes \bi r_B$, we have
\begin{eqnarray}
M' & = \beta' - \bi r'_A \otimes \bi r'_B  \\
& = O_A \beta O_B^T - O_A \bi r_A \otimes O_B \bi r_B \\
& =O_A \left( \beta - \bi r_A \otimes \bi r_B \right) O_B^T = O_A M O_B^T  .
\end{eqnarray}

The rank of $M$ is defined by the image of the linear map $f:\bi x \mapsto M \bi x$:
\begin{eqnarray}
\mathrm{rk}(M) = \dim \mathrm{Image}(f) = \dim \lbrace \bi y \in \mathbb{R}^{d_A} : \bi y = M \bi x, \, \bi x \in \mathbb{R}^{d_B} \rbrace  .
\end{eqnarray}
Since $O_A$ and $O_B$ are orthogonal matrices, we have
\begin{eqnarray}
\fl \lbrace \bi y \in \mathbb{R}^{d_A} : \bi y  = M' \bi x, \, \bi x \in \mathbb{R}^{d_B} \rbrace & = & \lbrace \bi y \in \mathbb{R}^{d_A} : \bi y = O_A M O_B^T \bi x, \, \bi x \in \mathbb{R}^{d_B} \rbrace \\
& = &  \lbrace \bi y \in \mathbb{R}^{d_B} : (O_A^T \bi y) = M (O_B^T \bi x), \, \bi x \in \mathbb{R}^{d_B} \rbrace  .
\end{eqnarray}
Since the maps $\bi y \mapsto \bi z =O_A^T \bi y $ and $\bi x \mapsto \bi w =O_B^T \bi x $ are bijective, we finally obtain
\begin{eqnarray}
\mathrm{rk}(M') = \dim \lbrace \bi z \in \mathbb{R}^{d_A} : \bi z = M \bi w, \, \bi w \in \mathbb{R}^{d_B} \rbrace = \mathrm{rk}(M).
\end{eqnarray}

In the following, we analyse the impact of the collective dephasing process, to be introduced in the next section, on the different concepts that were introduced in this section, all of which are intimately related to correlations in quantum states.

\section{Collective dephasing: Kraus map representation}
\subsection{Ensemble average dynamics}
Collective dephasing describes the dynamics of $N$ particles that share the same environment, but experience no direct particle-particle interactions among each other. The environment does not induce exchange of energy with the particles, and hence does not lead to dissipation, but rather to pure dephasing, i.e., loss of phase relations without loss of populations. A physically relevant example of such a scenario was already mentioned in the introduction: when an ensemble of atomic dipoles is spatially confined in a region where the electromagnetic field is homogeneous, all dipoles share the same transition frequency. The fluctuations of the field then lead to a collective dephasing process \cite{gross10,monz13}. The quantum state that predicts the measurement results after many experimental repetitions is described by the ensemble average over the actual realisations of these fluctuations \cite{ben,letter}.

Let us consider magnetic dipoles, described by two-level systems, in a constant magnetic field $\bi B$. The Hamiltonian $H$ of the $N$-particle system is given by $H=\gamma \bi B \cdot \bi S$, with $\bi S=\sum_{i=1}^N \boldsymbol{\sigma}^{(i)}$, where here $\boldsymbol{\sigma}^{(i)}$ is a vector of Pauli matrices on the Hilbert space of the $i$th particle, and further constants are absorbed into $\gamma$. Identifying $\hbar\omega/2=\gamma B$, with $\bi B = B \bi n$, we write
\begin{eqnarray}
H=\frac{\hbar\omega}{2} \sum_{i=1}^N \bi n\cdot \boldsymbol{\sigma}^{(i)}.
\end{eqnarray}
For a fixed magnetic field strength $B$, the fully coherent dynamics of the total system is given by
\begin{eqnarray}
\rho(t)=\rme^{-\rmi \omega t\bi n\cdot\boldsymbol{\sigma}/2}\otimes\cdots\otimes \rme^{-\rmi \omega t\bi n\cdot\boldsymbol{\sigma}/2} \rho(0)\rme^{\rmi\omega t\bi n\cdot\boldsymbol{\sigma}/2}\otimes\cdots\otimes \rme^{\rmi \omega t\bi n\cdot\boldsymbol{\sigma}/2},
\end{eqnarray}
where $\rho(t)$ is the $N$-particle density operator at time $t$. The dephasing process is now described by an ensemble average over the fluctuations of $\bi B$. The physically intuitive reason for such a description is the necessity to repeat experiments many times to produce significant statistics for the efficient estimation of the populations. We make the following assumptions on the fluctuations of $\bi B$:
\begin{itemize}
\item the \textit{direction} $\bi n$ of the magnetic field is constant, and the fluctuations only affect the field strength $B$;
\item the magnetic field may change from experimental run to the next, but within each run we assume the magnetic field to be constant.
\end{itemize}
Both of these assumptions can be motivated at the hand of state-of-the-art experiments on cold atoms or trapped ions: the external field influences the energy splitting of the atomic two-level systems through a Zeeman effect, as described above, and, thus, the field is chosen strong enough to dominate over the effect of possible stray fields. The field therefore has a fixed direction (satisfying the first of the two above assumptions), but small fluctuations of the supplying currents will produce weak intensity fluctuations of $B$, on top of a relatively large mean value. The mean value determines the time scale of the atomic evolution, which is therefore much faster than the time scale on which fluctuations occur (satisfying the second assumption).

Characterizing the intensity fluctuations with the probability distribution $p(\omega)$, the collective dephasing dynamics is described by
\begin{eqnarray}\label{eq.integral}
\fl \rho(t)=\int p(\omega) \rme^{-\rmi \omega t\bi n\cdot\boldsymbol{\sigma}/2}\otimes\cdots\otimes \rme^{-\rmi \omega t\bi n\cdot\boldsymbol{\sigma}/2} \rho(0)\rme^{\rmi \omega t\bi n\cdot\boldsymbol{\sigma}/2}\otimes\cdots\otimes \rme^{\rmi \omega t\bi n\cdot\boldsymbol{\sigma}/2} \rmd \omega .
\end{eqnarray}
In \cite{letter}, the above integral was solved analytically without further assumptions, and in the following we will recall the resulting solution and some of its properties.

\subsection{General properties}
The transient time evolution of a quantum state under collective dephasing (\ref{eq.integral}) is determined by the characteristic function
\begin{eqnarray}
\varphi(t)=\int d \omega p(\omega)e^{i\omega t}
\end{eqnarray}
of the probability distribution $p(\omega)$. Introducing the matrix elements
\begin{eqnarray}
M_{ij}(t) = \varphi\left[(i-j)t \right]
\end{eqnarray}
and the Hermitian operators
\begin{eqnarray}\label{eq.krausops}
\Theta_j = \frac{1}{j! (N-j)!} \sum_{s \in \Sigma_N} V_s\left[\Lambda_-^{\otimes j} \otimes \Lambda_+^{\otimes N-j} \right]V_s^\dagger  ,
\end{eqnarray}
where $\Lambda_\pm = \frac{1}{2}(\mathbb{I}_2 \pm \bi n \cdot \boldsymbol{\sigma})$, and $V_s =\sum_{i_1\dots i_N}|i_{s(1)}\dots i_{s(N)}\rangle\langle i_1\dots
i_N|$ represents the permutation $s$ in the Hilbert space of $N$ qubits, we can express the collective dephasing dynamics with the following map \cite{letter}:
\begin{eqnarray}\label{eq.colldeph}
\epsilon^{\bi n}_{t,0} : \rho(0) \rightarrow \rho(t) = \sum_{i,j=0}^N M_{ij}(t) \Theta_i \rho(0) \Theta_j.
\end{eqnarray}
The matrix of elements $M_{ij}$ is positive semi-definite, and can be diagonalized to obtain the canonical Kraus form of the above map \cite{letter}. One can further show \cite{letter} that the map~(\ref{eq.colldeph}) is always completely positive and trace preserving \cite{nielsen,geometry}. The obtained dynamics therefore exhibits the properties of dynamical maps associated with the dynamics of open quantum systems, indicating the equivalence of ensemble average approaches with open-system treatments based on a microscopic model for the environment and its coupling to the system \cite{breuerbook,GB13}.

In the context of the present article, we are only interested in the asymptotic limit, which is described, \textit{independently} of $p(\omega)$  (assuming that $p(\omega)$ is absolutely integrable), by \cite{letter}
\begin{eqnarray}\label{eq:solution-stationary}
\epsilon^{\bi n} : \rho(0) \rightarrow \rho_s = \lim_{t\rightarrow \infty} \rho(t) = \sum_{j=0}^N \Theta_j \rho(0) \Theta_j.
\end{eqnarray}
In performing this limit, we assume that the time evolution of the atomic ensemble is recorded for an interval long enough that the atomic evolution has reached its stationary state, but not too long to compromise with the assumption that the field strength can be considered constant during the evolution.

\subsection{Integral of motion}\label{ssec:conserved-trace}
In \cite{letter} the map (\ref{eq.colldeph}) was shown to conserve the trace of the $\beta$ matrix for bipartite systems, as defined in (\ref{eq:Fano-form}). This integral of motion can, in fact, be understood as the manifestation of the more general conservation of angular momentum in the special case of $N=2$. To see this, recall that the total spin $\bi S$ commutes with the Hamiltonian for every choice of the magnetic field $\bi B$, hence the expectation value of $\bi S^2=\bi S\cdot \bi S$ is conserved, even in the presence of the ensemble average over the fluctuations of the magnetic field.

We express the squared total spin as
\begin{eqnarray}\label{eq:total-spin}
\bi S^2 = \frac{\hbar^2}{4} \sum_{i=1}^N \boldsymbol{\sigma}^{(i)}\cdot\boldsymbol{\sigma}^{(i)} + \frac{\hbar^2}{2} \sum_{i,j=1 \atop i>j}^N \boldsymbol{\sigma}^{(i)} \cdot \boldsymbol{\sigma}^{(j)}  ,
\end{eqnarray}
with $\boldsymbol{\sigma}^{(i)} \cdot \boldsymbol{\sigma}^{(j)}= \sum_{k=1}^3 \sigma_k^{(i)}  \sigma_k^{(j)}$, and the index $k$ labels the spatial directions. We generalize the definition of the $\beta$ matrix to
\begin{eqnarray}\label{eq.generalbeta}
\beta_{ab}(t) = \tr \left\{ \rho(t) \cdot \sum_{i>j=1}^N \sigma_a^{(i)} \otimes\sigma_b^{(j)}  \right\}.
\end{eqnarray}
Note that this definition reduces to the bipartite $\beta$ matrix, as introduced in (\ref{eq:Fano-form}), in the special case of $N=2$. The total angular momentum is expressed via the quantum mechanical expectation value, using Eqs.~(\ref{eq:total-spin}) and (\ref{eq.generalbeta}),
\begin{eqnarray}
\langle \bi S^2 \rangle= \tr \{\rho(t)\bi S^2 \}= \frac{3 \hbar^2}{4} N + \frac{\hbar^2}{2}  \tr \beta(t),
\end{eqnarray}
and from the time-independence of $\langle \bi S^2 \rangle$, we obtain the conservation of the trace of the generalized $\beta$ matrix:
\begin{eqnarray}\label{eq.betaconserved}
\frac{\rmd}{\rmd t}\tr \beta (t)=0.
\end{eqnarray}

\subsection{Asymptotic collective dephasing of two qubits}\label{sec:application}
Let us discuss the description of the collective dephasing of an initial two-qubit state $\rho$ into the stationary state $\rho_s$, using the map~(\ref{eq:solution-stationary}) for $N=2$:
\begin{eqnarray}\label{eq:stat-bipartite}
\rho_s = \epsilon^{\bi n}[\rho] = \sum_{i=0}^2 \Theta_i \rho \Theta_i.
\end{eqnarray}
Based on (\ref{eq.krausops}), the Kraus operators $\Theta_i$ can be explicitly given as \cite{ben}
\begin{eqnarray}
 \Theta_0 &=\Lambda_+ \otimes \Lambda_+ = \frac{1}{4} \left(\mathbb{I}_2 \otimes \mathbb{I}_2 + \mathbb{I}_2 \otimes \bi n \cdot \boldsymbol{\sigma} + \bi n \cdot \boldsymbol{\sigma} \otimes \mathbb{I}_2 + \bi n \cdot \boldsymbol{\sigma} \otimes \bi n \cdot \boldsymbol{\sigma} \right), \\
 \Theta_1 &= \Lambda_+ \otimes \Lambda_- + \Lambda_- \otimes \Lambda_+ =\frac{1}{2} \left(\mathbb{I}_2 \otimes \mathbb{I}_2 - \bi n \cdot \boldsymbol{\sigma} \otimes \bi n \cdot \boldsymbol{\sigma} \right),\\
 \Theta_2 &= \Lambda_- \otimes \Lambda_- = \frac{1}{4} \left(\mathbb{I}_2 \otimes \mathbb{I}_2 - \mathbb{I}_2 \otimes \bi n \cdot \boldsymbol{\sigma} - \bi n \cdot \boldsymbol{\sigma} \otimes \mathbb{I}_2 + \bi n \cdot \boldsymbol{\sigma} \otimes \bi n \cdot \boldsymbol{\sigma} \right)  .
\end{eqnarray}

To efficiently describe the impact of the collective dephasing on an arbitrary initial state $\rho \doteq (\bi r_A, \bi r_B, \beta)$, we now derive a description of its map on the level of the reduced Bloch vectors $\bi r_A$ and $\bi r_B$, together with the $\beta$ matrix. We first express the $\beta$-matrix of the initial state in terms of a diagonal, singular value decomposition [recall (\ref{eq.transform})], as
\begin{eqnarray} \label{eq:svd}
\beta = \sum_{i=1}^3 d_i \bi v_i \otimes  \bi w_i  ,
\end{eqnarray}
where $\bi v_i $ and $ \bi w_i $ are normalized vectors, $d_i$ are non-negative real numbers, and the tensor product is defined element-wise as $(\bi a \otimes \bi b)_{kl} = a_k b_l $.

By direct application of the operators $\Theta_i$ and of the properties of the scalar product, (\ref{eq:stat-bipartite}) leads to \cite{ben}
\begin{eqnarray}
\epsilon^{\bi n} (\mathbb{I}_2 \otimes\mathbb{I}_2) &= \mathbb{I}_2\otimes\mathbb{I}_2,\\
\epsilon^{\bi n} (\mathbb{I}_2 \otimes \bi r \cdot \boldsymbol{\sigma} )&= \mathbb{I}_2 \otimes (\bi r \cdot \bi n) \bi n \cdot \boldsymbol{\sigma},\\
\epsilon^{\bi n} ( \bi r \cdot \boldsymbol{\sigma} \otimes \mathbb{I}_2)&= (\bi r \cdot \bi n) \bi n \cdot \boldsymbol{\sigma} \otimes \mathbb{I}_2,\\
\epsilon^{\bi n}(\bi v \cdot \boldsymbol{\sigma} \otimes  \bi w \cdot \boldsymbol{\sigma}) & =\frac{1}{2} \lbrace 2 (\bi n \cdot \bi v) \bi n \cdot \boldsymbol{\sigma} \otimes (\bi n \cdot \bi w) \bi n \cdot \boldsymbol{\sigma}  + (\bi v \times \bi n) \cdot \boldsymbol{\sigma} \otimes (\bi w \times \bi n) \cdot \boldsymbol{\sigma} \nonumber \\
& \qquad +  [\bi v - (\bi v \cdot \bi n) \bi n]\cdot \boldsymbol{\sigma} \otimes [\bi w - (\bi w \cdot \bi n) \bi n]\cdot \boldsymbol{\sigma} \rbrace  .
\end{eqnarray}
We can thus formulate \emph{transformation rules} for vectors, and tensor products thereof, to express how they are altered by the collective dephasing, as a function of the direction $\bi n$ of the fluctuating, external field:
\begin{eqnarray}
& \bi r \overset{\epsilon^{\bi n}}{\longrightarrow} (\bi r \cdot \bi n)\bi n \label{eq:corr-deph-bloch}\\
& \bi v \otimes  \bi w  \overset{\epsilon^{\bi n}}{\longrightarrow} \frac{1}{2} \lbrace 2 (\bi n \cdot \bi v) \bi n   \otimes (\bi n \cdot \bi w) \bi n    + (\bi v \times \bi n)   \otimes (\bi w \times \bi n)   \nonumber \\
& \qquad \qquad \qquad +  [\bi v - (\bi v \cdot \bi n) \bi n]  \otimes [\bi w - (\bi w \cdot \bi n) \bi n]  \rbrace. \label{eq:corr-deph-beta}
\end{eqnarray}
The local vectors that determine the decomposition on the right-hand-side of (\ref{eq:corr-deph-beta}) form an orthogonal basis consisting (in the first subsystem) of the direction $\bi n$ of the magnetic field, the vector orthogonal to the plane spanned by $ \bi n$ and $ \bi v $, and the vector orthogonal to these; the same holds for the second subsystem when $\bi v$ is replaced by $\bi w$. The decisive parameters are the angles $\cos\theta_{\bi v}=\bi n\cdot\bi v/\|\bi v\|$ and $\cos\theta_{\bi w}=\bi n\cdot\bi w/\|\bi w\|$.

Notice that if either one of $\bi v$ and $\bi w$ is parallel or orthogonal to $\bi n$, some of the terms in (\ref{eq:corr-deph-beta}) disappear. For now, we assume that $0 < |\cos \theta_{\bi v}| < 1$ and $0 < |\cos \theta_{\bi w}| < 1$. In this case, we can introduce two orthonormal bases of $\mathbb{R}^3$ as
\begin{eqnarray} \eqalign
\lbrace \bi s_1,\bi s_2,\bi s_3 \rbrace &=& \lbrace \frac{\bi v - (\bi v \cdot \bi n) \bi n}{\sin \theta_{\bi v}} , \bi n, \frac{(\bi v \times \bi n)}{\sin \theta_{\bi v}} \rbrace,\\
\lbrace \bi t_1,\bi t_2,\bi t_3 \rbrace &=& \lbrace \frac{\bi w - (\bi w \cdot \bi n) \bi n}{\sin \theta_{\bi w}} , \bi n, \frac{(\bi w \times \bi n)}{\sin \theta_{\bi w}} \rbrace   ,\label{eq:basis}
\end{eqnarray}
allowing us to re-express (\ref{eq:corr-deph-beta}) as
\begin{eqnarray}
\bi v \otimes \bi w \overset{\epsilon^{\bi n}}{\longrightarrow} \sum_{i=1}^3 \frac{\alpha_i}{2} \bi s_i \otimes \bi t_i,
\end{eqnarray}
with the coefficients
\begin{eqnarray}\label{eq:coeff}
\{\alpha_1,\alpha_2,\alpha_3\} = \lbrace \sin \theta_{\bi v} \sin \theta_{\bi w}, 2 \cos \theta_{\bi v}\cos \theta_{\bi w} , \sin \theta_{\bi v}\sin \theta_{\bi w} \rbrace.
\end{eqnarray}

\section{Generating correlations by collective dephasing}\label{sec:generate}
In this section we investigate to what extent the collective dephasing map (\ref{eq:solution-stationary}) can generate or increase the correlations between the subsystems. The present section extends the analysis of few special cases provided in the theoretical treatment of the experiment reported in \cite{ben} to a complete picture. 

We know the map is separable, hence it cannot create entanglement; however, it contains stochasticity, and therefore can create classical correlations between subsystems, thereby increasing the correlation rank $L$ of the initial state. For this reason, we focus on the analysis of the correlations in the asymptotic state, based on the correlation rank. As discussed in section~\ref{sec.correlationsintro}, a correlation rank of $L > d_{\min}$ can be interpreted as a witness for non-zero discord. Furthermore, we can exclude that the thereby detected discordant states can be generated by applying a local operation to a zero-discord state \cite{Gessner:2012fk}. For two-level systems we have $d_1 = d_2 = 2$, and, thus, the maximal correlation rank for zero-discord states is $L=d_{\min}=2$. States with $L=3$ or $L=4$ are considered strongly correlated, since their correlations are beyond the reach of any zero-discord states, and neither can be attained by states whose discord was generated by a local operation.

\subsection{Initially uncorrelated states}\label{sec:prod-state}
We begin by considering an initially completely uncorrelated state, i.e., a product state $\rho_{0} = \rho_A \otimes \rho_B$. In the Fano form (\ref{eq:Fano-form}) this reads
\begin{eqnarray}
\rho_0 = \frac{1}{2} \left( \mathbb{I}_2 + \bi r_A \cdot \boldsymbol{\sigma} \right) \otimes \frac{1}{2} \left( \mathbb{I}_2 + \bi r_B \cdot \boldsymbol{\sigma} \right) \doteq (\bi r_A, \bi r_B, \bi r_A \otimes \bi r_B)  .
\end{eqnarray}
This means that the initial $\beta$ matrix is $\beta_0 = \bi r_A \otimes \bi r_B$ and the initial rank  (\ref{eq:corr-rank}) is $L_0 = 1$, which is consistent with the state having no correlations \cite{ben}.

Application of the collective dephasing map, assuming that $\bi n $ does not coincide with the direction of the Bloch vectors $\bi r_A$ and $\bi r_B$ of the respective reduced systems, yields
\begin{eqnarray}\label{eq.asymptoticstate}
\rho_1  = \epsilon^{\bi n}(\rho_0) \doteq \left((\bi r_A \cdot \bi n) \bi n, \; (\bi r_B \cdot \bi n) \bi n, \; \sum_{i=1}^3 \frac{d_i}{2} \bi v_i \otimes  \bi w_i  \right)  ,
\end{eqnarray}
with the set of orthonormal vectors (\ref{eq:basis})
\begin{eqnarray}\label{eq:basisr} \eqalign
\lbrace \bi v_1,\bi v_2,\bi v_3 \rbrace &= \lbrace \frac{\bi r_A - (\bi r_A \cdot \bi n) \bi n}{r_A \sin \theta_A}, \bi n, \frac{\bi r_A \times \bi n}{r_A \sin \theta_A} \rbrace ,\\
\lbrace \bi w_1,\bi w_2,\bi w_3 \rbrace &= \lbrace \frac{\bi r_B - (\bi r_B \cdot \bi n) \bi n}{r_B \sin \theta_B}, \bi n, \frac{\bi r_B \times \bi n}{r_B \sin \theta_B} \rbrace  ,
\end{eqnarray}
and coefficients (\ref{eq:coeff})
\begin{eqnarray}
d_i \in \left\{r_A r_B \sin \theta_A \sin \theta_B, 2 r_A r_B \cos \theta_A \cos \theta_B, r_A r_B \sin \theta_A \sin \theta_B \right\}.
\end{eqnarray}
The set of accessible final states is described by four real parameters, namely the norms $r_{A,B}=\|\bi r_{A,B}\|$ of the reduced Bloch vectors and their angles $\theta_{A,B}$ with the magnetic field direction. Within the set of density matrices, which is a fifteen-dimensional real space, this represents a measure-zero set, therefore we cannot synthesize arbitrary states by adjusting the parameters of the collective dephasing map or of the initial state. In the following, however, we specify conditions that lead to a given value of the correlation rank $L$ after collective dephasing.

As discussed before, the vectors $\{\bi v_1,\bi v_2,\bi v_3\}$, defined in (\ref{eq:basisr}), span an orthonormal basis of $\mathbb{R}^3$. Following Sec.~\ref{sec:unitary-equivalence}, we can apply a local unitary operator $U_B$ (associated with a rotation matrix $O_B$, such that $O_B \bi w_i = \bi v_i $) which does not change the properties of interest. We apply this rotation only to the second subsystem, as described by the unitary operator $U=\mathbb{I}_A \otimes U_B$. This transforms the correlation matrix $\beta_1$ of the asymptotic state~(\ref{eq.asymptoticstate}) to the matrix $\beta'_1\sim\beta_1$, where $\beta'_1$, expressed in the basis $\{\bi v_1,\bi v_2,\bi v_3\}$, reads
\begin{eqnarray}\label{eq.diagonalbeta}
\beta'_1 = \frac{r_A r_B}{2} \pmatrix{
\sin \theta_A \sin \theta_B& 0&0 \cr
0&2 \cos \theta_A \cos \theta_B & 0 \cr
0&0& \sin \theta_A \sin \theta_B}.
\end{eqnarray}
When expressed in the same basis, the Bloch vectors after dephasing are given by
\begin{eqnarray}
\bi r_{A,B} \overset{\epsilon^{\bi n}}{\longrightarrow} \left( \bi r_{A,B} \cdot \bi n \right) \bi n = \left(0, r_{A,B}\cos \theta_{A,B}, 0 \right)^T  .
\end{eqnarray}
The final correlation rank reads, using (\ref{eq:corr-rank}),
\begin{eqnarray}
L_1 &=& 1 + \mathrm{rk} \left( \beta'_1 - r_A \cos \theta_A  \bi n \otimes r_B \cos \theta_B \bi n \right) \\
 & =& 1 + \mathrm{rk}  \pmatrix{
\sin \theta_A \sin \theta_B  & 0&0 \cr
0& 0 & 0\cr
0&0&  \sin \theta_A \sin \theta_B  \cr
}.
\end{eqnarray}

The correlation rank of the final state can only have two values, determined by the relative orientations of the reduced Bloch vectors and the direction of the magnetic field \cite{ben}:
\begin{itemize}
\item $L_1 = 1$, if the magnetic field is parallel to one of the reduced Bloch vectors: $\bi r_{A,B} \parallel \bi n$;
\item $L_1 = 3$, if the magnetic field has a different direction than both reduced Bloch vectors: $\bi r_{A,B} \nparallel \bi n$.
\end{itemize}

The physical interpretation is immediate if we realise that, when the Bloch vector of a subsystem coincides with the magnetic field direction, that subsystem is in an eigenstate of the local Hamiltonian $\hbar\omega\bi{n}\cdot\boldsymbol{\sigma}/2$, and is consequently invariant under the action of the map. In this case, this atom can be treated separately, and the collective dephasing acts only on the remaining atoms, whose Bloch vector differs from $\bi n$. Let us suppose that $\bi n = \bi r_A / r_A$: the atom described by $\mathcal{H}_A$ is no longer affected by the collective dephasing process and the final state can be written as
\begin{eqnarray}
\rho_1 = \epsilon^{\bi n} \left( \rho_A \otimes \rho_B \right) = \rho_A \otimes\epsilon^{\bi n} \left( \rho_B \right),
\end{eqnarray}
which is an uncorrelated product state with correlation rank $L=1$. As is best seen from the integral representation~(\ref{eq.integral}), this observation is easily generalized for systems of $N>2$ atoms. In general, for product states involving arbitrary local states $\rho^{(i)}$ of qubits $i$ that satisfy
\begin{eqnarray}\label{eq.factorizationcondition}
[\bi n\cdot\boldsymbol{\sigma}^{(i)},\rho^{(i)}]=0,
\end{eqnarray}
the dephasing operation factorizes:
\begin{eqnarray}\label{eq.factorization}
\epsilon^{\bi{n}}(\rho\otimes\rho^{(i)})=\epsilon^{\bi{n}}(\rho)\otimes\rho^{(i)},
\end{eqnarray}
where $\rho$ is an arbitrary quantum state of the remaining qubits. The states that satisfy the factorisation condition~(\ref{eq.factorizationcondition}) encompass all incoherent mixtures of eigenstates of $ \bi n\cdot\boldsymbol{\sigma}^{(i)}$ and the identity operator.

In short, application of (\ref{eq:solution-stationary}) to an uncorrelated state produces a state with high correlation rank, as long as $\bi r_{A,B} \nparallel \bi n$, but cannot reach the maximal value of $L=4$.
Yet, is it possible to transform the resulting $L=3$ state into an $L=4$ state by a consecutive, second application of the collective dephasing map? Since we consider collective dephasing to the asymptotic state, the second application would not have any effect unless we \textit{change} the direction of the external field. This can be shown as follows: the application of the collective dephasing map along the direction $\bi m \neq \bi n \equiv \bi e_2$ yields a state with a $\beta$ matrix
\begin{equation*}
\fl \beta_2 \sim \cos \theta_A \cos \theta_B \epsilon^{\bi m} (\bi e_2 \otimes \bi e_2)
 + \frac{\sin \theta_A \sin \theta_B}{2} [\epsilon^{\bi m}(\bi e_1 \otimes \bi e_1) + \epsilon^{\bi m}( \bi e_3 \otimes \bi e_3)] .
\end{equation*}
If we define $q = \cot \theta_A \cot \theta_B$, the correlation rank of the state is given by $L_2 = 1 + \mathrm{rk} M$, where
\begin{eqnarray}
M & = \frac{\beta_2}{\sin \theta_A \sin \theta_B} - q (\bi e_2 \cdot \bi m)^2 \bi m \otimes \bi m \\
 & = q [\epsilon^{\bi m}(\bi e_2 \otimes \bi e_2) - (\bi e_2 \cdot \bi m)^2 \bi m \otimes \bi m )] + \epsilon^{\bi m}(\bi e_1 \otimes \bi e_1) + \epsilon^{\bi m}( \bi e_3 \otimes \bi e_3) 
\end{eqnarray}
where $\sin \theta_A \sin \theta_B$ is non-zero if $\bi r_{A,B} \nparallel \bi n$. We can now observe that, when $\bi m = \bi e_2$, the term proportional to $q$ vanishes and $M = \bi e_1 \otimes \bi e_1 + \bi e_3 \otimes \bi e_3$, yielding $L_2 = 3$; alternatively, we can solve the equation $\epsilon^{\bi m}(\bi e_1 \otimes \bi e_1) + \epsilon^{\bi m}( \bi e_3 \otimes \bi e_3)  = 0$ with the constraint $\Vert \bi m \Vert = 1$ to obtain $M = (q/2) [(\bi e_2 - (\bi e_2 \cdot \bi m) \bi m ) \otimes (\bi e_2 - (\bi e_2 \cdot \bi m) \bi m ) + (\bi e_2 \times \bi m) \otimes (\bi e_2 \times \bi m)]$, which yields again $L_2 = 3$. Finally, one can show that the matrix $M$ is not of full rank, e.g. by solving $\det M = 0 $ as a function of $\bi m$ given the parameter $q$. In any case, using linear algebra or by direct computation, one can verify that there exists at most a measure-zero set of directions, other than the one of the first dephasing process, along which the correlation rank will not increase. Therefore, an $L=1$ state can be converted into an $L=4$ state by a twofold application of the collective dephasing map, if the direction of the magnetic field is different for the second dephasing.

\subsection{Initial states of correlation rank $L=2$}\label{sec:class-corr}
After studying the influence of the collective dephasing on an initially uncorrelated state $L_0=1$, we now turn to the discussion of an initial state with low correlation rank $L_0 = 2$. This state is considered weakly correlated since it contains correlations that are compatible with either a state of zero discord, or a non-zero discord state that can be created from a state of zero discord with a local operation---recall the discussion at the end of section~\ref{sec.correlationsintro}. We consider states with maximally mixed reduced subsystems, i.e., systems with vanishing reduced Bloch vectors, which can be written as $\rho_0\doteq (0,0,\beta_0)$. Since these states are diagonal in the basis of Bell states, they are also called \emph{Bell-diagonal} states. In the Fano form (\ref{eq:Fano-form}) they are written as \cite{horodecki96}
\begin{eqnarray}\label{eq:class-corr-state}
\rho_0 = \frac{1}{4} \left(\mathbb{I}_4 + d \: \bi v \cdot \boldsymbol{\sigma} \otimes \bi w \cdot \boldsymbol{\sigma} \right),
\end{eqnarray}
where $d\neq 0$ and the initial $\beta$ matrix was expressed in terms of its singular value decomposition (\ref{eq:svd}), as $\beta_0 = d \bi v \otimes \bi w $. Positivity of the state $\rho_0$ requires that $\vert d \vert \leq 1$. States of this form have always zero discord \cite{ben}, as can be easily seen based on the cummutativity of the local operators \cite{PhysRevLett.105.190502}, as discussed in Sec.~\ref{sec.correlationsintro}.

Application of the collective dephasing map (\ref{eq:corr-deph-beta}) produces another Bell-diagonal state, $\rho_1\doteq(0,0,\beta_1)$, where the correlation matrix $\beta_1$ can be written (analogously to the case discussed before (\ref{eq.diagonalbeta}), and possibly after suitable, local orthogonal transformations that do not alter the correlation properties) as
\begin{eqnarray}
\beta_1 \sim \frac{d}{2} \pmatrix{
 \sin \theta_{\bi v} \sin \theta_{\bi w} &0&0 \cr
0&2 \cos \theta_{\bi v} \cos \theta_{\bi w}&0 \cr
0&0&  \sin \theta_{\bi v} \sin \theta_{\bi w}  \cr
}.
\end{eqnarray}

In order to find the correlation rank of the final state, by virtue of (\ref{eq:corr-rank}), we need to determine $L_1 = 1 + \mathrm{rk}\left(\beta_1 \right)$. Except for the trivial case $d = 0$, we notice that also in this case the rank depends on the geometric features of the state, namely on the angle between the magnetic field direction and the left- and right-singular vectors of $\beta_0$: the state has correlation rank \cite{ben}
\begin{itemize}
\item $L_1 = 1$ if either ($\bi v \parallel \bi n $ and $\bi w \perp \bi n$), or ($\bi w \parallel \bi n $ and $\bi v \perp \bi n$);
\item $L_1 = 2$ if $\bi v \parallel \bi n $ or $\bi w \parallel \bi n $ (but neither $\bi v \perp \bi n$ nor $\bi w \perp \bi n$);
\item $L_1 = 3$ if $\bi v \perp \bi n $ or $\bi w \perp \bi n $ (but neither $\bi v \parallel \bi n $ nor $\bi w \parallel \bi n $);
\item $L_1 = 4$ otherwise.
\end{itemize}
A weakly correlated ($L_0=2$) Bell-diagonal state can therefore be transformed into a state with maximal correlation rank ($L=4$), provided that the magnetic field direction does not coincide with some very specific choices, determined by the geometric characterisation of the initial state.

\subsection{Applications}\label{sec.applications}
\textbf{Generation of Werner states.}---An important class of states for many applications of quantum information theory is given by the family of Werner states,
\begin{eqnarray}\label{eq:werner-state}
\rho_W = s \ket{\Psi^-}\bra{\Psi^-} + (1-s) \frac{\mathbb{I}_4}{4}  ,
\end{eqnarray}
with the singlet state  $\ket{\Psi^-} =(\ket 0 \ket 1 - \ket 1 \ket 0)/\sqrt{2}$, and $-\frac{1}{3} < s \leq 1$ \cite{werner}. Written in the Fano form (\ref{eq:Fano-form}), this state has $\beta = - \mathrm{diag}(s,s,s)$.

\begin{figure}
\centering
\begin{subfigure}{0.48\textwidth}
\includegraphics[width=\textwidth]{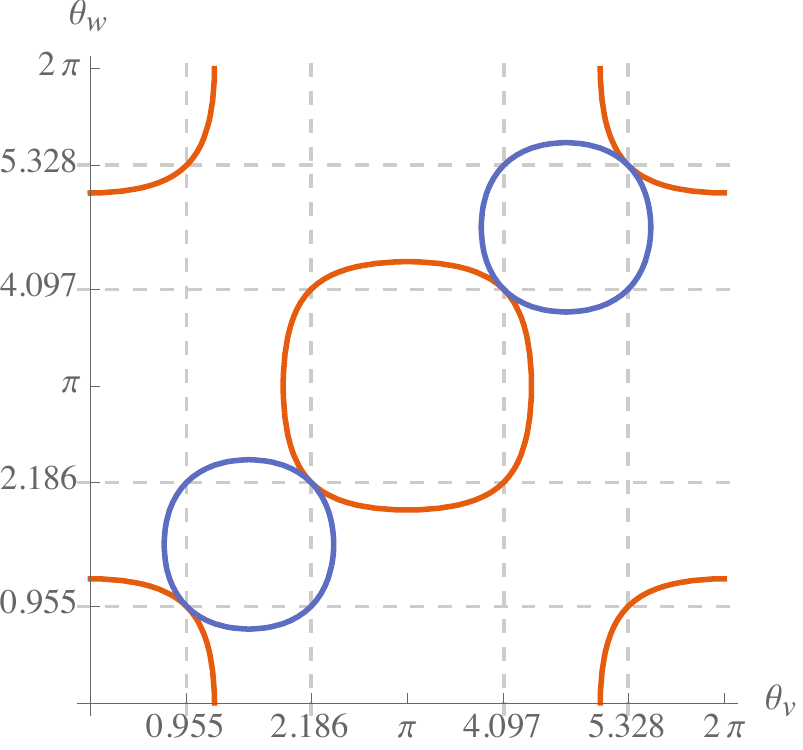}
\end{subfigure}
\begin{subfigure}{0.48\textwidth}
\includegraphics[width=\textwidth]{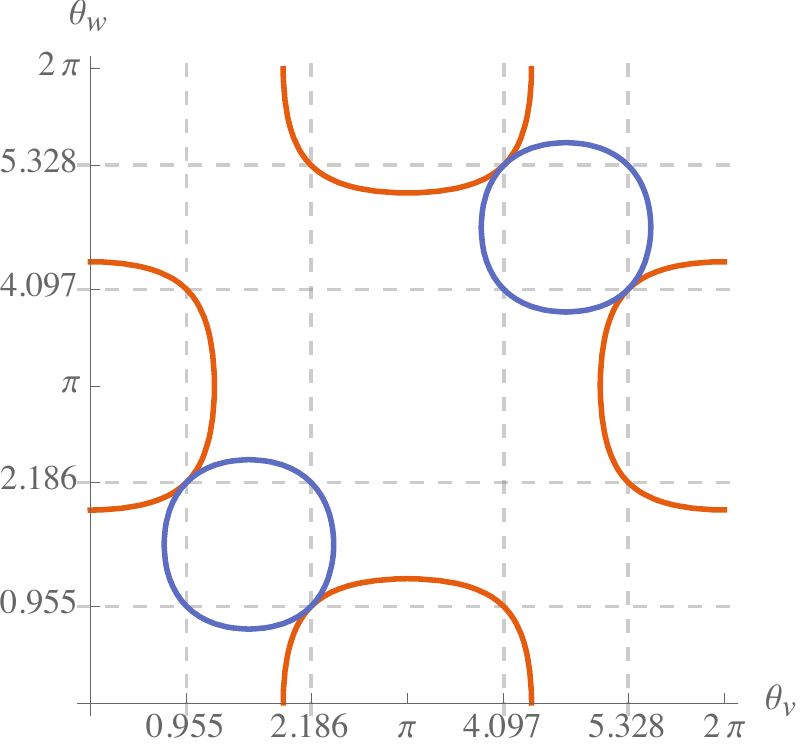}
\end{subfigure}
\caption{Pairs of angles that solve the system (\ref{eq:wern-state-sys}) for the Werner state (\ref{eq:werner-state}) (left), and the system (\ref{eq:general-werner-sys}) for the Werner-like state (\ref{eq:werner-like}) (right), in the case $k=1/3$. The blue (darker) circles are the contour lines of $ \sin \theta_{\bi v} \sin \theta_{\bi w}  = 2/3$ [first equation in (\ref{eq:wern-state-sys}) and (\ref{eq:general-werner-sys})], the orange (lighter) ones, those of $\cos \theta_{\bi v} \cos \theta_{\bi w} = \pm 1/3$ (second equation). The solutions of the two systems (\ref{eq:wern-state-sys}) and (\ref{eq:general-werner-sys}) are then the intersections of the curves.}
\label{fig:werner-minus}
\end{figure}

Collective dephasing can generate such a state from an $L=2$ Bell-diagonal state of the form~(\ref{eq:class-corr-state}) when we choose the values for $d$, $\theta_{\bi v}$ and $\theta_{\bi w}$ that solve the following system of equations:
\begin{eqnarray}\label{eq:wern-state-sys}
\cases{\sin \theta_{\bi v} \sin \theta_{\bi w}  = 2 k \\ \cos \theta_{\bi v} \cos \theta_{\bi w} = k}  ,
\end{eqnarray}
where $k = - s/d$ and $d \neq 0$. Summation of the two equations yields
\begin{eqnarray}
\cos \theta_{\bi v} \cos \theta_{\bi w}  + \sin \theta_{\bi v} \sin \theta_{\bi w} = \cos(\theta_{\bi v} - \theta_{\bi w}) = 3k  ,
\end{eqnarray}
which proves that the system of equations~(\ref{eq:wern-state-sys}) admits solutions in the variables $\{\theta_{\bi v}, \theta_{\bi w} \}$ only if $\vert k \vert \leq 1/3$. In particular, when $k =1/3$, the solutions lie on the lines $\theta_{\bi w} = \theta_{\bi v}$ (see figure \ref{fig:werner-minus}, left). The solutions are then found as:
\begin{eqnarray}
\theta_{\bi v} =\theta_{\bi w} = \pm \arcsin\left(\sqrt{2/3} \right) \approx \pm 0.955,
\end{eqnarray}
or
\begin{eqnarray}
\theta_{\bi v} =\theta_{\bi w} = \pi\pm \arcsin\left(\sqrt{2/3} \right).
\end{eqnarray}

We remark here that we have to simultaneously respect the conditions $\vert k \vert \leq 1/3$ and $\vert d \vert \leq 1$ (for positivity of the state), which leads to the observation that, based on the present approach, it is impossible to generate Werner states with $s > 1/3$. In fact, the Werner states are separable precisely when $s\leq1/3$ and entangled when $s>1/3$. This is consistent with the fact that the initial state was separable and that the map cannot create entanglement.

The same procedure can be followed to produce Werner-like states
\begin{eqnarray}\label{eq:werner-like}
\rho_\alpha= s \ket{\alpha}\bra{\alpha} + (1-s) \frac{\mathbb{I}_4}{4}  ,
\end{eqnarray}
where $\ket{\alpha}$ is one of the other Bell states:
\begin{eqnarray}
\eqalign{
\ket{\Psi^+} & = \frac{1}{\sqrt{2}}\left(\ket 0 \ket 1 + \ket 1 \ket 0\right) \\
\ket{\Phi^+} & = \frac{1}{\sqrt{2}} \left(\ket 0 \ket 0 + \ket 1 \ket 1 \right) \\
\ket{\Phi^-} & = \frac{1}{\sqrt{2}} \left(\ket 0 \ket 0 - \ket 1 \ket 1 \right)
.}
\end{eqnarray}
These states  have $\beta$ matrices with the eigenvalues \cite{horodecki96}:
\begin{eqnarray} \eqalign{
\ket{\Psi^+}\bra{\Psi^+} & \doteq(0,0,\mathrm{diag}(s,s,-s)), \\
\ket{\Phi^+}\bra{\Phi^+} &\doteq(0,0,\mathrm{diag}(s,-s,s)), \\
\ket{\Phi^-}\bra{\Phi^-} &\doteq (0,0,\mathrm{diag}(-s,s,s)).}
\end{eqnarray}
The values of $\theta_{\bi v}$ and $\theta_{\bi w}$ that yield those states are found by solving the system
\begin{eqnarray}\label{eq:general-werner-sys}
\cases{\sin \theta_{\bi v} \sin \theta_{\bi w}  = 2 k \\ \cos \theta_{\bi v} \cos \theta_{\bi w} = - k}  .
\end{eqnarray}

In fact, we will show in the next section that the $\beta$ matrix of the asymptotic state always has one non-degenerate eigenvalue, with the associated eigenspace spanned by $\bi n$, and one doubly degenerate eigenvalue, with the plane orthogonal to $\bi n$ as eigenspace. Thus, the three Werner-like states are distinguished by the choice of the direction $\bi n$ of the magnetic field. In all previous expressions we had arbitrarily chosen to identify $\bi n$ with the $y$-direction, $\bi n =\bi e_2$, thus the non-degenerate eigenvalue always appeared as  second diagonal element, corresponding to $\ket{\Phi^+}$. In general, we identify
\begin{eqnarray}\label{eq:state-direc-rel}\eqalign{
 \ket{\Phi^-}  & \quad \bi n \equiv \bi e_1\\
 \ket{\Phi^+} \quad \iff & \quad  \bi n \equiv \bi e_2 \\
\ket{\Psi^+} & \quad \bi n \equiv \bi e_3 .}
\end{eqnarray}
In order to generate these Werner-like states we then solve the system of equations (\ref{eq:general-werner-sys}) to find the relative angles between $\bi n$, $\bi v$ and $\bi w$, and rotate the basis accordingly, to obtain the target state. In order to solve (\ref{eq:general-werner-sys}), the same conditions on $k$ hold, but the solutions for $k = 1/3$  now lie on the lines $\theta_{\bi w} = \pi - \theta_{\bi v}$  (see figure \ref{fig:werner-minus}, right), with $\theta_{\bi v}$ as before: $\theta_{\bi v} = \pm \arcsin\left(\sqrt{2/3} \right)$ or $ \theta_{\bi v} = \pi \pm \arcsin\left(\sqrt{2/3} \right)$. For the cases with $k \neq 1/3$ there will be twice as many solutions, which can be easily found numerically.

For the genuine (singlet-based) Werner state it does not matter how we choose to order the eigenvalues of the $\beta$ matrix (by appropriate choice of the coordinate system relative to the magnetic field orientation), since they are all negative. This expresses these states' invariance under arbitrary collective unitary rotations of the form $U\otimes U$, which effectively rotate the coordinate system. In general, the set of $U\otimes\cdots\otimes U$-invariant states can be identified as a completely time-invariant set of states under the collective, multipartite dephasing operation, for arbitrary directions of the magnetic field \cite{letter}.

\textbf{Generating resource states for entanglement distribution.}---Separable states can find applications in quantum information protocols such as \textit{entanglement distribution} \cite{Cubitt,PhysRevLett.108.250501,Chuan:2012kx}; for experimental realisations based on separable states see \cite{ex3,ex1,ex2}. This protocol allows to increase the entanglement between two parties by exchanging a carrier particle, which is not necessarily entangled with the two parties. It is, however, necessary that the combined state contains non-zero discord between the two parties and the carrier. In fact it was recently further shown that discord alone is not sufficient, and that discordant mixtures of two pure product states are not able to distribute entanglement \cite{Strel14}. By producing states with $L>2$, we can ensure that these cannot be written as mixtures of two product states. 

Let us consider the following example of an initial two-qubit state \cite{Chuan:2012kx}
\begin{eqnarray}
\rho_{AB} = p \ket{\psi_s} \bra{\psi_s} +\frac{1-p}{4} \mathbb{I}_4,
\end{eqnarray}
with $\ket{\psi_s} = \sqrt{s} \ket{0} \ket{0} + \sqrt{1-s} \ket{1} \ket{1}$. When $s=1/2$, these are the Werner-like states, (\ref{eq:werner-like}), generated by $\ket{\Phi^+}$; in particular, the protocol can be realised with an initially separable state, which further imposes $p \leq 1/3$. As explained in the previous paragraph, these states can be generated from a weakly correlated ($L=2$) state by action of the collective dephasing map. States with other values of $s$ are not accessible as the map only produces states with a $\beta$ matrix that contains a doubly-degenerate eigenvalue. In the three-partite scenario considered here, the Werner-like state above can be generated on systems $AB$ if the initial state contains no correlations with system $C$, which is prepared in a maximally mixed state $\rho_C = \mathbb{I}_2/2$. By virtue of~(\ref{eq.factorization}) the qubit $C$ is invariant under the collective dephasing dynamics. The state $\rho_{AB}\otimes\rho_C$ can thus be generated by collective dephasing of a weakly correlated $L=2$ state of zero discord for the choice of parameters $s = 1/2$ and, e.g., $p = 1/3$. For these parameters the same state was shown to be useful for entanglement distribution \cite{Chuan:2012kx}.

\textbf{Entanglement activation from uncorrelated states through collective dephasing.}---A three-partite scenario is also considered for the \textit{entanglement activation} protocol \cite{PhysRevLett.106.160401,Piani2011}, where initial discord between systems $A$ and $B$ is converted into entanglement across the partition $AB|M$, where $M$ is an initially independent ancilla system that interacts via a local unitary operation with system $B$. Formally, this can be interpreted as a local measurement process of the system $B$, where $M$ is a measurement apparatus. Thereby, the entanglement of the combined quantum state describing the apparatus and the system before readout (state reduction) of the apparatus is then directly linked to the discord of the initial bipartite state of $A$ and $B$ \cite{PhysRevLett.106.160401}.

An all-optical experimental realisation of this protocol confirmed this general theoretical result by modelling all three subsystems $A$, $B$ and $M$, as polarization qubits \cite{Adesso}. Based on the collective dephasing process, we can now extend this protocol such that also initial zero-discord states or even completely uncorrelated states can be used to generate entanglement. To this end, we subject an initial three-qubit state of the form
\begin{eqnarray}\label{eq.intentactstate}
\rho_0=\rho_{AB}\otimes |1\rangle\langle 1|_M
\end{eqnarray}
to a three-partite collective dephasing process in direction $\bi n$, where we choose $|1\rangle$ as an eigenstate of the operator $\bi n\cdot \boldsymbol{\sigma}$, and $\rho_{AB}$ may be an arbitrary state of qubits $A$ and $B$. While in the previous section we have made use of the local invariance under the dephasing process, see (\ref{eq.factorization}), for the trivial case of the identity operator, here we prepare the third qubit in an eigenstate of the local Hamiltonian to achieve the same effect. Application of the collective dephasing map to the state~(\ref{eq.intentactstate}) yields
\begin{eqnarray}
\epsilon^{\bi n}(\rho_0)=\epsilon^{\bi n}(\rho_{AB})\otimes |1\rangle\langle 1|_M.
\end{eqnarray}
By following the conditions provided in sections \ref{sec:prod-state} and \ref{sec:class-corr}, initial product states of Bell-diagonal zero discord states $\rho_{AB}$ can be easily converted into strongly correlated, non-zero discord states $\epsilon^{\bi n}(\rho_{AB})$. Based on the entanglement activation protocol, any local unitary operation on $BM$ necessarily must generate distillable entanglement across the $AB|M$ partition, and the amount of generated entanglement is lower bounded by the discord of the state $\epsilon^{\bi n}(\rho_{AB})$ \cite{PhysRevLett.106.160401,Piani2011}. In the present section we only considered the asymptotic mapping of the collective dephasing map (\ref{eq:solution-stationary}) after long times, but the results hold also for intermediate times $t$, as described by the map (\ref{eq.colldeph}).

In a recent experiment \cite{Orieux}, \textit{local} noise processes were used to generate discordant states (see also \cite{ben}), whose discord was subsequently activated into entanglement.

\textbf{Local quantum interferometry.}---To end this section, we briefly comment on the use of correlated separable states in the context of local precision measurements \cite{Girolami}. The field of quantum metrology is dedicated to developing methods that allow to estimate unknown parameters with the highest possible precision, often by exploiting the usage of entangled states \cite{Giovanetti,Smerzi,Review}. Consider a setup, in which two incoming particles are sent into two different arms of an interferometer before being jointly measured. The parameter to be estimated is a local phase shift $\varphi$, imprinted on one of the particles by an unknown Hamiltonian $H^{(i)}$ (with fixed, non-degenerate spectrum) through the unitary operation $U^{(i)}_{\varphi}=e^{-i\varphi H^{(i)}}$. In a worst-case scenario, the local Hamiltonian may commute with the quantum state, which therefore renders any estimation of the phase shift impossible. This, however, is only possible if the quantum state has zero discord. In general, the worst-case estimation precision of $\varphi$ is quantified by a particular measure of discord \cite{Girolami}.

Consequently, the successful estimation of the phase in the above scenario requires the presence of non-zero discord between the two particles. Using the results provided in sections \ref{sec:prod-state} and \ref{sec:class-corr}, the required discord can be easily generated by submitting the two parties to a collective dephasing process \textit{before} sending them into the interferometer.

\subsection{Summary}
In the present section, we have discussed the behaviour of the correlation rank under the action of the collective dephasing dynamics. We have seen in section~\ref{sec:prod-state} that completely uncorrelated initial states ($L=1$) can be transformed into states with $L=3$ by a single application of collective dephasing, or into states of $L=4$ by a double application, provided that the magnetic field direction is changed before the second dephasing. Weakly correlated states with $L=2$ are transformed into states with the maximum correlation rank $L=4$ for most choices of the magnetic field direction, as discussed in section~\ref{sec:class-corr}. Since any value of $L\geq 3$ implies the presence of nonzero discord, the strongly correlated states that can be conveniently generated by the collective dephasing process allow for direct applications in a series of tasks from quantum information theory, as shown in section~\ref{sec.applications}.

\section{Protecting correlations under collective dephasing}\label{sec:generalk}
\subsection{Initial states of arbitrary correlation rank}\label{ssec:eigenvalues}
In the previous section we saw how the collective nature of the ensemble averaged dephasing process, induced by a spatially homogeneous, fluctuating external field, can be used to generate strongly correlated quantum states, which have direct applications for specific tasks in quantum information processing.

We now extend the discussion from the initially weakly correlated states to initial states with an arbitrary correlation rank $L_0 \leq 4$. We thus shift our focus from the generation of strongly correlated quantum states, to the robustness of the correlations under the action of collective dephasing. The objective, then, is to control the magnetic field orientation, such that the preservation of these correlations under the collective dephasing is ensured.

As in the previous section, we assume Bell-diagonal states (thus vanishing reduced Bloch vectors $\bi r_{A,B}$). The matrices $\beta$ and $\beta - \bi r_A \otimes \bi r_B$ have the same rank unless $r_A r_B$ is a singular value of $\beta$ with left- and right-singular vectors $\bi r_A / r_A$ and $\bi r_B / r_B$, respectively, and only in these cases can the correlation rank be reduced by non-zero reduced Bloch vectors [recall (\ref{eq:corr-rank})]. The effect of the collective dephasing map on arbitrary initial states is easily investigated based on (\ref{eq:stat-bipartite}).

Furthermore, before application of the map (\ref{eq:corr-deph-beta}), we employ unitary transformations to bring the $\beta$ matrix of the initial state into diagonal form $\beta_0 = \mathrm{diag}\left(d_1, d_2, d_3 \right)$. Its three eigenvalues then parametrize \cite{horodecki96} the Fano form (\ref{eq:Fano-form}) of the initial state:
\begin{eqnarray}\label{eq:bd-state}
\rho_0 = \frac{1}{4} \left(\mathbb{I}_4 + \sum_{i=1}^3	d_i \bi e_i \cdot \boldsymbol{\sigma} \otimes \bi e_i \cdot \boldsymbol{\sigma}	\right)  ,
\end{eqnarray}
where $\lbrace \bi e_i \rbrace_i$ is the standard basis in $\mathbb{R}^3$. This transformation neither affects the correlation rank of the initial state, as discussed earlier, nor that of the final state, unless we start from a rank-2 state of the form (\ref{eq:class-corr-state}) and we apply a magnetic field in the direction $\bi v$ or $\bi w$, as dealt with in Sec. \ref{sec:class-corr}. In all other cases, there is at least one rank-1 matrix of the form $\bi v_i \otimes \bi w_i$ in (\ref{eq:svd}) that is transformed, according to (\ref{eq:corr-deph-beta}), into a rank-3 matrix by the collective dephasing map. 

Application of the collective dephasing map (\ref{eq:corr-deph-beta}) yields 
\begin{eqnarray}\label{eq:beta-renorm}
\beta_0 \overset{\epsilon^{\bi n}}{\longrightarrow}
\beta_1 = \sum_{i=1}^3 \frac{d_i}{2}\lbrace (1-n_i^2) \left[ \bi a_i \otimes\bi a_i + \bi b_i  \otimes\bi b_i \right] + 2 n_i^2 \bi n \otimes \bi n  \rbrace,
\end{eqnarray}
with the normalized vectors
\begin{eqnarray}
\bi a_i = \frac{\bi e_i - n_i \bi n}{\sqrt{1-n_i^2}}, \quad \bi b_i = \frac{\bi e_i \times \bi n}{\sqrt{1-n_i^2}}.
\end{eqnarray}
Notice that all the $\bi a_i$'s and $\bi b_i$'s belong to the plane orthogonal to $\bi n$ and are mutually orthogonal: $\bi a_i \cdot \bi b_i = 0, \forall i$. This means that we may write all orthonormal bases $\lbrace \bi a_i, \bi b_i \rbrace$, with $i=1 \ldots 3$, as a rotation about $\bi n$  of, e.g., $\lbrace \bi a_1, \bi b_1 \rbrace$ by an angle $\varphi_i$ (in this case $\varphi_1 = 0$):
\begin{eqnarray}
& \bi a_i = \cos \varphi_i \, \bi a_1 +  \sin \varphi_i \, \bi b_1 \\
& \bi b_i = - \sin \varphi_i \, \bi a_1 +  \cos \varphi_i \, \bi b_1  .
\end{eqnarray}
By direct substitution we obtain
\begin{eqnarray}
\bi a_i \otimes \bi a_i + \bi b_i \otimes \bi b_i =
\left[ \cos^2 \varphi_i + \sin^2\varphi_i \right] (\bi a_1 \otimes\bi a_1 + \bi b_1  \otimes\bi b_1) ,
\end{eqnarray}
which shows that $ \bi a_i \otimes\bi a_i + \bi b_i  \otimes\bi b_i =  \bi a_1 \otimes\bi a_1 + \bi b_1  \otimes\bi b_1, \forall i$. The $\beta$ matrix of the final state, given by (\ref{eq:beta-renorm}), can thus be rewritten as
\begin{eqnarray}
\beta_1 = \sum_{i=1}^3 \frac{d_i}{2} \lbrace (1-n_i^2) \left[\bi a_1 \otimes\bi a_1 + \bi b_1  \otimes \bi b_1 \right] + 2 n_i^2 \bi n \otimes \bi n  \rbrace  .
\end{eqnarray}
Because $\lbrace \bi a_1, \bi n, \bi b_1 \rbrace$ is an orthonormal basis of $\mathbb{R}^3$, the above expression is a spectral decomposition of $\beta_1$ where two eigenvalues appear, one of which is two-fold degenerate:
\begin{eqnarray}
\lambda_1 (\bi n) &= \sum_{i=1}^3 d_i n_i^2 \label{eq:lambda1}, \\
 \lambda_2(\bi n) &= \frac{1}{2}\sum_{i=1}^3 d_i (1-n_i^2) . \label{eq:finallambda2}
\end{eqnarray}
From these expressions the invariance of the eigenvalues under cyclic permutation of the indices is evident. The $\beta$ matrix of the state after application of the map is given by
\begin{eqnarray}
\beta_1 = \lambda_1(\bi n)\, \bi n \otimes \bi n + \lambda_2(\bi n) \left[\bi a_1 \otimes \bi a_1 + \bi b_1 \otimes \bi b_1 \right]  ,
\end{eqnarray}
where we could have equivalently chosen $\lbrace \bi a_2, \bi b_2 \rbrace$ and $\lbrace \bi a_3, \bi b_3 \rbrace$, instead of $\lbrace \bi a_1, \bi b_1 \rbrace$ (we have used above their equivalence under rotation of the basis about $\bi n$, which is orthogonal to all of them).

To summarize, we find
\begin{eqnarray}\label{eq:beta-diag}
\beta_0 = \mathrm{diag}\left(d_1, d_2, d_3 \right) \overset{\epsilon^{\bi n}}{\longrightarrow}
\beta_1 \sim \mathrm{diag} \left( \lambda_2 (\bi n),  \lambda_1 (\bi n),  \lambda_2 (\bi n)\right).
\end{eqnarray}
From the above explicit expressions for the eigenvalues we can immediately verify that the trace of the $\beta$ matrix is preserved under the application of the map [recall the general result, (\ref{eq.betaconserved})]:
\begin{eqnarray}\label{eq:cons-tr}
\tr \beta_1 = \sum_{i=1}^3 d_i n_i^2 + 2 \times \frac{1}{2} \sum_{i=1}^3 d_i \left( 1-n_i^2 \right) = \sum_{i=1}^3 d_i = \tr \beta_0.
\end{eqnarray}

\subsection{Geometric description}\label{ssec:geom-descr}
Bell-diagonal states of two qubits allow for a simple geometric description of their correlations properties \cite{horodecki96}, which we will employ in the following. Recall that Bell-diagonal states are unambiguously characterized by their $\beta$ matrix (\ref{eq:bd-state}). Since there exists a unique unitary operator that diagonalizes this matrix without changing the state's correlation properties, we can parametrize the correlation properties of Bell-diagonal states by the three real eigenvalues of $\beta$. It follows that this unitary operator defines an isomorphism that maps each Bell-diagonal state to a point $\bi d = \left( d_1, d_2, d_3 \right) \in \mathbb{R}^3$. Positivity implies that the space of density matrices is isomorphic to a tetrahedron $\mathcal{T}$ whose vertices represent the four Bell states \cite{horodecki96}:
\begin{eqnarray}\eqalign{
 \ket{\Psi^-}\bra{\Psi^-} &\sim B_0 = (-1,-1,-1) \\
 \ket{\Phi^-}\bra{\Phi^-} &\sim B_1 = (-1,1,1)   \\
 \ket{\Phi^+}\bra{\Phi^+}  &\sim B_2 = (1,-1,1)  \\
 \ket{\Psi^+}\bra{\Psi^+} &\sim B_3 = (1,1,-1).} \label{eq:vertices}
\end{eqnarray}

In this tetrahedron we distinguish two regions: a central octahedron $\mathcal{O}$ with vertices $\pm \bi e_i,i=1\ldots 3$, which contains the separable states, and the four remaining corners, containing the entangled states. Because each corner has one of the Bell states as its vertex, we call the other entangled states in that corner ``Bell-like''. We notice that all $B_0$-like entangled states have negative coordinates, while the $B_i$-like states, with $i=1\ldots 3$, have only the $i$-th coordinate negative, like $B_i$ itself.

\begin{figure}
\centering
\includegraphics[scale=0.8]{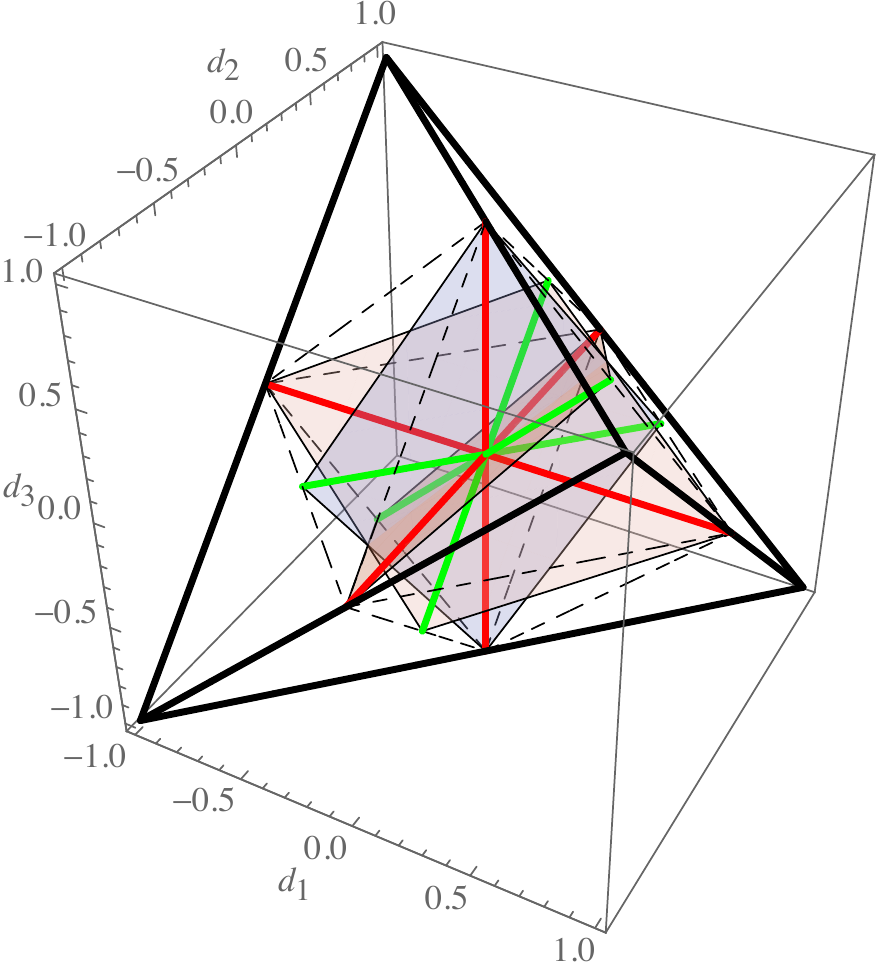}
\caption{Points of interest inside the tetrahedron of Bell-diagonal states: the red lines represent the rank-2 states (points with one non-zero coordinate), the planes represent the rank-3 states (points with one vanishing coordinate) and the green lines represent the rank-3 states that are reachable via the map (\ref{eq:stat-bipartite}) (points with one vanishing and two equal coordinates).}
\label{fig:tetra-poi}
\end{figure}

In figure \ref{fig:tetra-poi} we show how the classes of states we are interested in are represented in the tetrahedron $\mathcal{T}$. The origin is the maximally mixed state $\rho_m=\mathbb{I}_4/4$, and the rotation axes of the octahedron correspond to the rank-2 states. The rank-3 states lie on the squares obtained by intersecting the octahedron with the planes orthogonal to its rotation axes. Both these sets of states (rank-2 and 3) have measure zero inside the tetrahedron. Moreover, as discussed in Sec. \ref{sec:correlations}, all entangled states have rank 4, but the converse is not true.

\subsection{Geometric action of the map}\label{ssec:geometric-action}

We now turn to describing the action of the map (\ref{eq:corr-deph-beta}) in this geometrical framework. In particular, the subset of states that is accessible by the map is defined by the conservation of the trace of the $\beta$ matrix (Sec. \ref{ssec:conserved-trace}) and the double degeneracy in the coordinates of the asymptotic state (Sec. \ref{ssec:eigenvalues}).

\begin{figure}
\centering
\begin{subfigure}{0.45\textwidth}
\includegraphics[width=\textwidth]{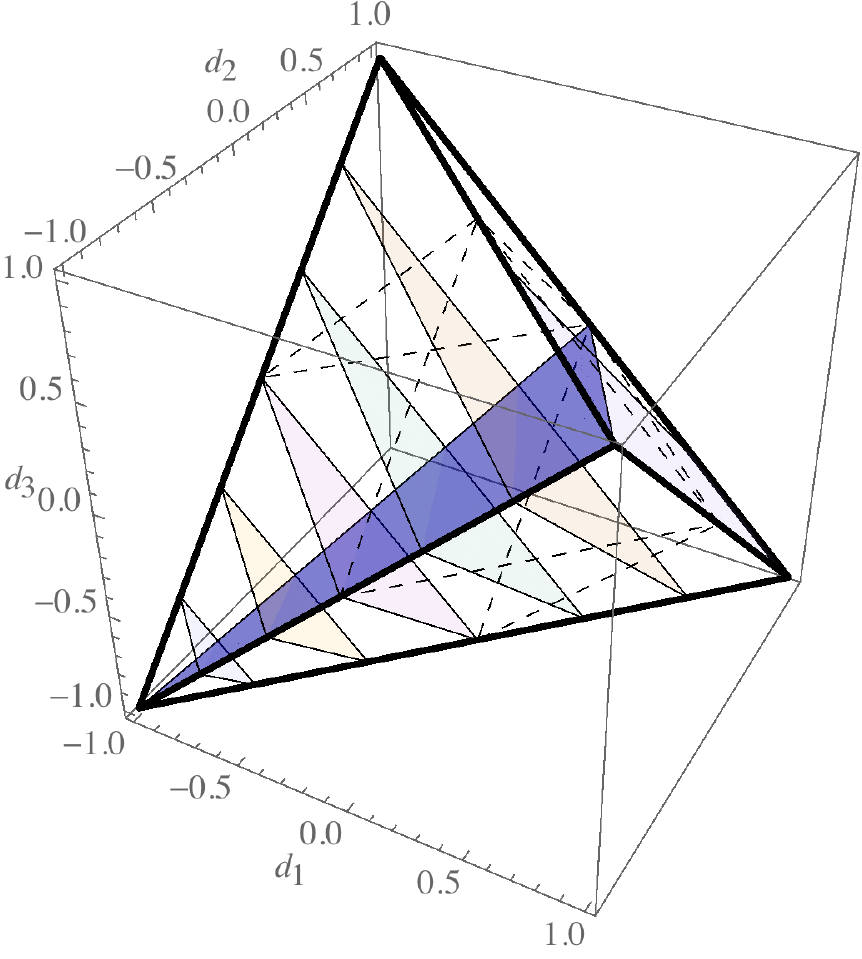}
\subcaption{The plane $\Pi_2$, (\ref{eq:pi-plane}), identifies states whose $\beta$ matrix has two equal eigenvalues (dark blue), and the family of parallel planes $\Gamma_k$, (\ref{eq:gamma-plane}), identifies constant values of $\tr \beta$ (lighter shades).}
\label{fig:par_planes}
\end{subfigure}
\hfill
\begin{subfigure}{0.45\textwidth}
\includegraphics[width=\textwidth]{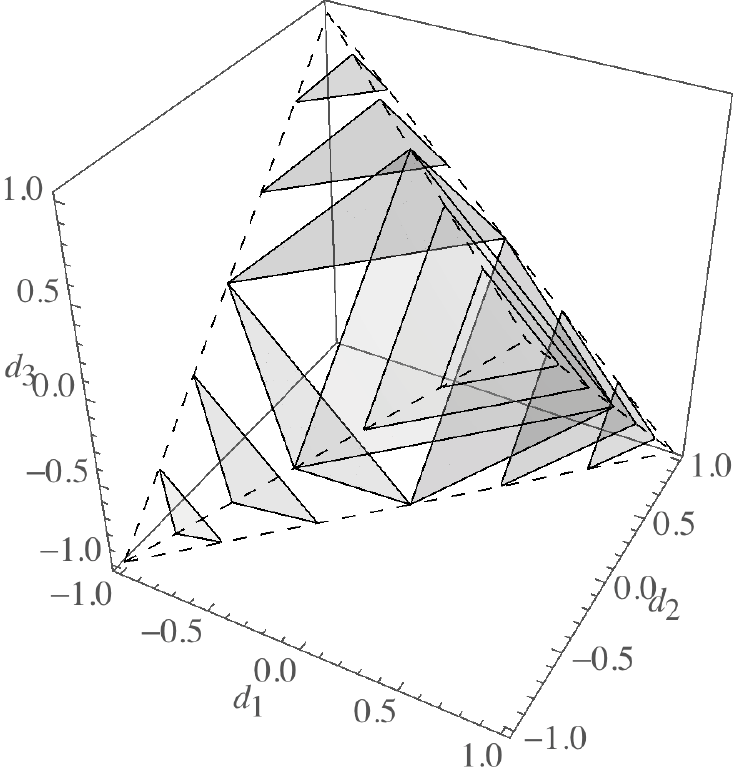}
\subcaption{``Iso-concurrence'' planes in the tetrahedron of Bell-diagonal states.}
\label{fig:isoconc}
\end{subfigure}
\caption{Geometric interpretation of the action of the collective dephasing map (\ref{eq:stat-bipartite}) on the Bell-diagonal states (\ref{eq:bd-state}) contained in the tetrahedron. The octahedron, inscribed into the tetrahedron, represents the separable states.}
\end{figure}

In the following we adopt Greek letters to indicate cyclic indices, i.e., $\lbrace \nu -1 , \nu, \nu+1 \rbrace$ denotes an even permutation of $\lbrace 1, 2 ,3 \rbrace$. Whenever we apply the map (\ref{eq:corr-deph-beta}) we transform to the reference frame where the direction of the magnetic field is the unit vector $\bi e_\nu$ of the standard basis. This allows for the most general description, which is independent of the choice of reference frame. Consequently, the set of points with two equal coordinates is a plane defined as
\begin{eqnarray}\label{eq:pi-plane}
\Pi_\nu = \lbrace \left(d_1,d_2,d_3 \right) \in \mathcal{T} : d_{\nu-1} = d_{\nu+1} \rbrace,
\end{eqnarray}
while the conservation of $\tr \beta$ defines another plane
\begin{eqnarray}\label{eq:gamma-plane}
\Gamma_k = \lbrace \left(d_1 , d_2, d_3 \right) \in \mathcal{T} : d_1 + d_2 + d_3 = k \rbrace  ,
\end{eqnarray}
where, inside the tetrahedron, $-3 \leq k \leq 1$. Because the trace of the $\beta$ matrix is conserved at all times, the trajectory of each point lies on the $\Gamma_k$ plane defined by the initial coordinates. However, since the final state must belong to $\Pi_\nu$, every state will asymptotically move to the intersection line defined by $\Gamma_k \cap \Pi_\nu$, whose existence is guaranteed by the fact that $\Pi_\nu \perp \Gamma_k, \forall k,\nu$  (figure \ref{fig:par_planes}).

The position of the final state on the line defined by $\Gamma_k \cap \Pi_\nu$ is determined by the magnetic field direction. Let us use again $\lambda_1(\bi n)$ and $\lambda_2(\bi n) = (k-\lambda_1(\bi n))/2$ as the coordinates of the final states. We obtain
\begin{eqnarray}
\beta_1 \sim \lambda_1(\bi n) \bi e_\nu + \frac{k-\lambda_1(\bi n)}{2} (\bi e_{\nu-1}+\bi e_{\nu+1}).
\end{eqnarray}
Since $k$ is fixed by the initial state, the coordinates depend only on $\bi n$.

We remark here that, when the initial state is $B_0$-like, the intersection lines never cross the octahedron of separable states, because the planes $\Gamma_k$ are parallel to the octahedron face opposite to $B_0$ (figure \ref{fig:par_planes}). We deduce that the states in the $B_0$ corner move entirely in that corner. On the other hand, the entangled states in the other corners move inside their respective corner, but may also enter the octahedron. This entails significant implications for the entanglement preservation of the initial states from the different corners, and ultimately enables the effect of \textit{time-invariant} entanglement for $B_0$-like states, as we will discuss in further detail later in this manuscript.

\subsection{Loss of coherence}
In the geometric representation of Bell-diagonal states, planes of equal purity are characterized by constant values of
\begin{eqnarray}\label{eq.purity}
\mathcal{P}(\bi d)=\frac{1}{4}(1+d_1^2+d_2^2+d_3^2),
\end{eqnarray}
which correspond to concentric spheres around the tetrahedron's center, the maximally mixed state. 

From the Kraus representation~(\ref{eq.colldeph}) of the collective dephasing map $\epsilon^{\bi n}_{t,0}$, it follows immediately that the purity of the final state is upper bounded, for all $t\geq 0$, by the initial state's purity:
\begin{eqnarray}
\tr \{\rho(t)\rho(t)\}\leq \tr \{\rho(0)\rho(0)\}.
\end{eqnarray}
In most cases, however, the collective dephasing tends to push states towards the center of the tetrahedron. The preservation of the initial purity can be achieved for time-invariant states. Whether there exist states outside of invariant subspaces for which the purity is preserved is unclear, but seems unlikely since the entire dynamics must lie on the circumference generated by the intersection of the sphere defined by $\mathcal{P}=\mathrm{const}$ and the plane $\Gamma_k$.

\subsection{Impact on the correlation rank and discord}\label{ssec:impact-corrank}
We now focus on the correlation rank of the final states, which directly translates into the number of non-vanishing coordinates of a point in the tetrahedron. 

The states in the $B_0$ corner never exhibit a vanishing coordinate, as displayed in figure \ref{fig:tetra-poi}. Moreover, as discussed in the previous section, the states in this corner never enter the octahedron, hence they never intersect the subsets of rank-2 and rank-3 states. We conclude that the correlation rank in the states in the $B_0$ corner is robust and always maximal, which also implies that these states are always discordant, and cannot be created using local operations on classical states.

For the remaining states, i.e., those in the octahedron and in the other three corners, there is at most one intersection between the reachable rank-2 or rank-3 states, and the accessible final states defined by the line $\Pi_\nu \cap \Gamma_k $, where $-1 \leq k \leq 1$ for those states. The magnetic field direction that yields such a state is found by solving either $\lambda_1(\bi n) = 0$ or $\lambda_2(\bi n) = 0$ under the normalisation constraint $\Vert \bi n \Vert =1$. Using the Kronecker-Capelli theorem \cite{shilov} it is possible to show that there exists at most a 1-parameter family of directions that solve such equations. We conclude that, after the action of the collective dephasing map (\ref{eq:corr-deph-beta}), the state has almost always a correlation rank equal to 4. This includes both scenarios where a state was already initially discordant and this property is preserved throughout the dephasing dynamics, as well as when the discord is generated by the action of the collective dephasing map, recall also the experiment reported in \cite{ben}.

\subsection{Protection of initial two-qubit entanglement}\label{ssec:ent-protection}
Let $\rho$ be a Bell-diagonal state of two qubits and $p_0 \ldots p_3$ its eigenvalues in the basis of Bell states. If $p_\mathrm{max} = \max(p_0 \ldots p_3)$, then the concurrence of $\rho$ can be written as \cite{0305-4470-34-47-329}
\begin{eqnarray}\label{eq:conc-bell-diag}
C(\rho) = \max(0,2p_\mathrm{max}-1)  .
\end{eqnarray}
Inside the tetrahedron $\mathcal{T}$, where states are described by vectors $\bi d \in \mathbb{R}^3$, the concurrence has an isotropic form,
\begin{eqnarray}\label{eq:iso-conc}
\mathcal{C}(\bi d) = \frac{1}{2} \max(0,\sum_i \vert d_i \vert -1)  ,
\end{eqnarray}
which can be interpreted as the distance to the faces of the octahedron of separable states, parametrized by the equation $\sum_i \vert d_i \vert = 1$. Following the state space from these faces to either one of the corners of the tetrahedron, the concurrence increases from zero to one.

Let us remind here that the collective dephasing map, being separable, cannot create entanglement. In our geometrical framework, this means that points initially in the octahedron $\mathcal{O}$ are mapped onto $\Pi_\nu \cap \mathcal{O}$. Suppose that an initial state is inside the octahedron, i.e., it has vanishing concurrence (\ref{eq:iso-conc}): $\sum_{i=1}^3 \vert d_i \vert \leq 1$. The  concurrence (\ref{eq:iso-conc}) in the final state, where the coordinates are $\lambda_{1(2)}(\bi n)$ from (\ref{eq:lambda1}) and (\ref{eq:finallambda2}), is then $ \vert \lambda_1(\bi n) \vert + 2 \vert \lambda_2(\bi n)\vert$ and we have the following chain of inequalities:
\begin{eqnarray}
\fl \left\vert \sum_{i=1}^3 d_i n_i^2\right\vert + \left\vert \sum_{i=1}^3 d_i (1- n_i^2)\right\vert \leq \sum_{i=1}^3 \vert d_i \vert n_i^2 +  \sum_{i=1}^3 \vert d_i \vert (1- n_i^2) = \sum_{i=1}^3 \vert d_i \vert \leq 1  ,
\end{eqnarray}
where we have used the normalisation condition $\Vert \bi n \Vert =1 $, which implies $n_i^2 \leq 1, \forall i$.

Since the states in the $\ket{\Psi^-}$-corner are constrained to move on planes parallel to the octahedron face opposite to $B_0$, these states remain on iso-concurrence planes at all times (compare figs.~\ref{fig:par_planes} and \ref{fig:isoconc}). More specifically, as shown in \cite{letter}, the points in this corner are characterized by negative coordinates: $d_i \leq 0, \forall i$. Hence, the concurrence (\ref{eq:iso-conc}) for these states can be rewritten as
\begin{eqnarray}
\mathcal{C}(\bi d) = -\frac{1}{2} - \frac{1}{2} \sum_i d_i = \frac{1}{2} (-1-k)   ,
\end{eqnarray}
where we have used $\tr \beta = \sum_i d_i = k$ and, for entangled states in this corner, $k<-1$.

In the other corners of the tetrahedron only one of the coordinates is negative, and we denote it with $\nu$: $d_\nu \leq 0$, where $\bi d$ represents the initial state. We then have
\begin{eqnarray}
\fl \mathcal{C}(\bi d) =\frac{1}{2}\left( -1+\sum_{i = \nu-1}^{\nu+1} \left| d_i \right| \right)=\frac{1}{2}\left( -1 - d_\nu + d_{\nu-1} + d_{\nu+1}\right) = \frac{1}{2}\left(-1 + k -2 d_\nu \right)  .
\end{eqnarray}
Let  $\bi d^\mathrm{f}$ represent the final, asymptotic state of the system after collective dephasing, and let us once again denote the negative components of $\bi d$ and $\bi d^\mathrm{f}$ with a subscript $\nu$, i.e., we have $d_\nu\leq 0$ and $d^\mathrm{f}_\nu \leq 0$, respectively. Since the eigenvalue $\lambda_2(\bi n) $ has a double degeneracy, but the points in the $B_1 \ldots B_3$ corners can only have a single negative component, the negative eigenvalue after collective dephasing must necessarily be
\begin{eqnarray}
d_\nu^\mathrm{f} = \lambda_1(\bi n)= \sum_{i = \nu-1}^{\nu+1} d_i n_i^2.
\end{eqnarray}
The concurrence of the final state then reads
\begin{eqnarray}\label{eq:conc-final} \eqalign{
\mathcal{C} \left(\bi d^\mathrm{f}\right) &= \frac{1}{2} \max 	\left\lbrace 0 , -1 + k - 2 d_\nu^\mathrm{f} \right\rbrace \\
& = \frac{1}{2} \max 	\left\lbrace 0 , -1 + \sum_{i = \nu-1}^{\nu+1} (1 - 2 n_i^2) d_i \right\rbrace .}
\end{eqnarray}
It follows that $\mathcal{C}(\bi d) = \mathcal{C}(\bi d^\mathrm{f}) \iff \bi n = \pm \bi e_\nu$, i.e., the initial concurrence of states in the three triplet corners can only be preserved if the magnetic field is chosen along the special direction that characterizes the respective corner. Moreover, (\ref{eq:conc-final}) allows for an estimation of the loss of concurrence due to small deviations from $\bi e_\nu$.

Let us conclude this section by highlighting the relationship between the different types of states and the respective magnetic field directions: the concurrence of a $B_\nu$-like state is \emph{conserved} under collective dephasing if and only if the magnetic field points in the $\bi e_\nu$ direction, while in $B_0$-like states the concurrence is \emph{always conserved}, \emph{independently} of the direction of the magnetic field. This means that, independently of the field direction, one quarter of the entangled Bell-diagonal states shows preserved concurrence, while an additional quarter of states that depend on the magnetic field direction shows the same effect.

Let us briefly remark on the applicability of the results of the present section to states that are not Bell-diagonal, i.e., which have non-vanishing reduced Bloch vectors. To this end, let us consider a state $\rho^\mathrm{G} \doteq (\bi r_A, \bi r_B, \beta^\mathrm{G})$ and the local operation $\tau$ such that $\rho^\mathrm{BD} = \tau[\rho^\mathrm{G}] \doteq (0,0,\beta^\mathrm{BD})$ \cite{Aravind19977}.
The operation that transforms a general two-qubit into a Bell-diagonal state, i.e., sets the reduced Bloch vectors to zero, is a local operation that cannot increase entanglement \cite{Aravind19977}, therefore $C(\rho^\mathrm{BD}_s) \leq C(\rho^\mathrm{G}_s)$, where the behaviour of the lower bound $C(\rho^\mathrm{BD}_s)$ under collective dephasing can be controlled, e.g., by manipulating the magnetic field direction as described above.

Despite the possibility to achieve conservation of entanglement in a Bell-diagonal state, the concurrence of the more general state can still decrease due to the time evolution of the reduced Bloch vectors. Perfect conservation of concurrence in a general bipartite state is consequently only possible when concurrence is preserved in the corresponding Bell-diagonal state, \emph{and} when additionally \emph{both} initial reduced Bloch vectors are parallel to the magnetic field direction, which is the only possible case where the reduced Bloch vectors of the initial and asymptotic state coincide, as per (\ref{eq:corr-deph-bloch}).

\subsection{Time-invariant states vs. time-invariant entanglement}\label{ssec:time-invariant}
Let us review the insights gained in the previous section on the entanglement dynamics under collective dephasing employing the geometric representation of the tetrahedron picture. The state $\ket{\Psi^-} $, represented as one of the corners of the tetrahedron, is completely unaffected by the collective dephasing process because it is an eigenstate of the Hamiltonian of the system for \emph{every} magnetic field direction. The state is therefore time-invariant. This is true also for all the Werner states $s \ket{\Psi^-}\bra{\Psi^-} + (1-s) \frac{\mathbb{I}_4}{4}$. In the geometric framework, these states constitute the rotation axis of the tetrahedron that passes through $B_0$ and the origin.

For the other Bell states, as well as the respective mixtures with the identity (Werner-like states), the previous remarks only hold provided the magnetic field is chosen according to the association rule (\ref{eq:state-direc-rel}), since in that case the respective Bell state becomes an eigenstate of the Hamiltonian. This explains and generalizes the observations reported in \cite{robustwerner}, where a dynamical evolution equivalent to the collective dephasing along the $z$ axis is applied to the Werner and Werner-like states. Those obtained from the $\ket{\Psi^\pm}$ states are then categorized as `robust', while those from $\ket{\Phi^\pm}$ are `fragile'. In fact, as we have shown above, only the Werner state (\ref{eq:werner-state}) is robust under collective dephasing in any field direction, while the other Werner-like (\ref{eq:werner-like}) states are robust only when a specific direction is chosen \cite{letter}.

The preservation of entanglement within decoherence-free subspaces \cite{Palma,Duan,Lidar} -- which in the present case, due to the absence of a Hamiltonian evolution, coincide with the above-mentioned time-invariant subspaces -- is a rather obvious phenomenon: if the state does not evolve in time, then its properties are naturally also conserved. Our analysis, however, points out the non-trivial possibility of time-invariant entanglement of states beyond time-invariant subspaces \cite{letter}, see also \cite{Karpat}. In fact, as pointed out above, all states in the $B_0$ corner remain at a level of constant concurrence, even if they do not belong to the axis of time-invariant (Werner) states.

\subsection{Summary}
In the present section we have studied the behavior of initially correlated quantum states under the influence of the collective dephasing evolution. To this end, we have extended the analysis of the correlation rank $L$ under collective dephasing to states of arbitrary initial correlation rank. Furthermore, we have presented a convenient geometrical setting to describe Bell-diagonal states and their evolution under collective dephasing. This geometrical interpretation allowed us to demonstrate that highly correlated ($L=4$) initial states retain their strong correlations, as quantified by $L$, under the collective dephasing dynamics for almost all choices of the magnetic field direction. 

Combining the a geometric description of the collective dephasing dynamics with a geometric interpretation of the concurrence -- an entanglement measure -- we have characterized a finite set of states showing time-invariant entanglement despite their incoherent evolution.

\subsection{Extension to the multipartite case}
Some features of bipartite states can be directly generalized to a multipartite setting. For example, the generalized Werner states, i.e., those states that are invariant under $U^{\otimes N}$-operations \cite{werner,eggeling}, are the fixed points of the collective dephasing map for any number of qubits \cite{letter}. In \cite{letter}, the decay and the time-invariant preservation of multipartite entanglement properties was also provided. A compelling explanation for the mechanism that enables this phenomenon in a multipartite case is, however, presently unavailable. In the following we allude to two possible approaches towards achieving this goal.

The conservation of $\tr \beta$ in bipartite states of qubits can be related to the overlap between the state of the system $\rho(t)$ and the singlet state $\ket{\Psi^-}\bra{\Psi^-}$ \cite{letter}. This is true for any state $\ket{\phi}$ that is an eigenstate of the collective dephasing Hamiltonian for an arbitrary choice of the magnetic field direction. These eigenstates are the multi-qubit singlet states \cite{entdetection}, which exist only for an even number $N$ of qubits, and are a family of $N!/[(N/2)!(N/2+1)!]$ linearly independent states. The overlap with each of these states is an integral of motion, which in the bipartite case reduces to the conservation of entanglement for all magnetic field directions.
It is however not clear whether the overlap with the \emph{multi}-qubit singlet states is related to entanglement, especially considering that we observe conserved entanglement properties for any -- and not just for an even -- number of qubits.

Another approach originates from the observation that the collective dephasing map describes a simultaneous rotation of all qubits, hence it has a set of fixed points, the rotation axis, and the (hyper-)planes orthogonal to it are mapped onto (a subset of) themselves. The fixed points of the map are the Werner states \cite{letter}, and, in the tetrahedron of Bell-diagonal states, the planes orthogonal to the rotation axis contain the states with the same amount of entanglement. In an analogous higher-dimensional picture, the set of states orthogonal to the family of multipartite Werner states may lead to an interesting set of integrals of motion.

\section{Conclusions}

To summarize, we provided a detailed analysis of the impact of a collective dephasing process on the correlation properties of bipartite states. Based on the Kraus representation of the dephasing process \cite{letter}, we provided conditions that enable the generation of states with high correlation rank and non-vanishing quantum discord from uncorrelated or only weakly correlated states.

Using an intuitive geometric representation of the state space and the collective dynamics, we investigated the entanglement dynamics under collective dephasing. For initially entangled states, we provided conditions that ensure the complete preservation of the entanglement for all times under the dephasing dynamics. Surprisingly, this is possible even for large families of states that do not belong to time-invariant subspaces, i.e., states that change in time due to the incoherent dynamics. While some of the results could be generalized straight-forwardly to multipartite scenarios, a compelling picture describing time-invariant multipartite entanglement remains to be conceived.

Since collective dephasing represents one of the dominant sources of error for many experiments with trapped atomic particles, we expect that the results derived in the present article can be readily harnessed in state-of-the-art setups with trapped ions \cite{ben,monz13} or ultracold atoms \cite{gross15}.

\textit{Note added.}---An experimental observation of time-invariant entanglement was reported in \cite{TIENT} after completion of this manuscript.

\ack
M. G. thanks the German National Academic Foundation for their support. This project has received funding from the European Union's Horizon 2020 research and innovation programme
under grant agreement No 641277.


\section*{References}
\begin{thebibliography}{10}

\bibitem{wineland98}
Wineland D~J, Monroe C, Itano W~M, Leibfried D, King B~E and Meekhof D~M 1998
  \href{http://dx.doi.org/10.6028/jres.103.019}{{\em J. Res. Natl. Inst. Stand. Technol.\/} {\bf 103} 259}.

\bibitem{morsch}
Morsch O and Oberthaler M 2006
\href{http://journals.aps.org/rmp/abstract/10.1103/RevModPhys.78.179}{\textit{Rev. Mod. Phys.} \textbf{78} 179}.

\bibitem{ciraczoller}
Cirac J and Zoller P 1995 \href{http://dx.doi.org/10.1103/PhysRevLett.74.4091}{{\em Phys. Rev. Lett.\/} {\bf 74} 4091}.

\bibitem{hartmut}
H\"{a}ffner H, Roos C~F and Blatt R 2008 \href{http://dx.doi.org/10.1016/j.physrep.2008.09.003}{{\em Phys. Rep.\/} {\bf 469} 155}.

\bibitem{gross10}  Gross C, Zibold T,  Nicklas E, Est\`{e}ve J and Oberthaler M K 2010
\href{http://www.nature.com/nature/journal/v464/n7292/full/nature08919.html}{\textit{Nature} \textbf{464} 1165}.

\bibitem{ben}
Lanyon B~P, Jurcevic P, Hempel C, Gessner M, Vedral V, Blatt R and Roos C~F
  2013 \href{http://link.aps.org/doi/10.1103/PhysRevLett.111.100504}{{\em Phys. Rev. Lett.\/} {\bf 111} 100504}.

\bibitem{monz13}
Schindler P \etal 2013
  \href{http://stacks.iop.org/1367-2630/15/i=12/a=123012}{{\em New J. Phys.\/} {\bf 15} 123012}.

\bibitem{gross15}
Fukuhara T, Hild S, Zeiher J, Schau\ss{} P, Bloch I, Endres M and Gross C 2015 \href{http://dx.doi.org/10.1103/PhysRevLett.115.035302}{{\em Phys. Rev. Lett.\/} {\bf 115} 35302}.

\bibitem{breuerbook} Breuer H P and Petruccione F 2007 {\em The theory of open quantum systems} (Oxford University Press, UK).

\bibitem{horodecki^4}
Horodecki R, Horodecki P, Horodecki M and Horodecki K 2009 \href{http://link.aps.org/doi/10.1103/RevModPhys.81.865}{{\em Rev. Mod.
  Phys.\/} {\bf 81} 865}.
  
\bibitem{nielsen}
Chuang I~L and Nielsen M~A 2000 {\em {Quantum Computation and Quantum
  Information}\/} (Cambridge University Press).  

\bibitem{Benatti} Benatti F, Floreanini R and Olivares S 2012 \href{http://www.sciencedirect.com/science/article/pii/S0375960112009619}{\textit{Phys. Lett. A} \textbf{376} 2951}.

\bibitem{GB13} Gessner M and Breuer H P 2013 \href{http://journals.aps.org/pre/abstract/10.1103/PhysRevE.87.042128}{\textit{Phys. Rev. E} \textbf{87} 42128}.

\bibitem{matteos} Rossi M A C, Benedetti C and Paris M G A 2014 \href{http://www.worldscientific.com/doi/abs/10.1142/S0219749915600035}{\textit{Int. J. Quantum Inform.} \textbf{12} 1560003}.

\bibitem{letter}
Carnio E~G, Buchleitner A and Gessner M 2015  \href{http://journals.aps.org/prl/abstract/10.1103/PhysRevLett.115.010404}{{\em Phys. Rev. Lett.\/} {\bf 115} 10404}.

\bibitem{PhDGessner} Gessner M 2015 \textit{Dynamics and Characterization of Composite Quantum Systems} \href{http://www.freidok.uni-freiburg.de/data/10214}{PhD Thesis, Albert-Ludwigs-Universit\"at Freiburg}.

\bibitem{Gessner:2012fk}
Gessner M, Laine E~M, Breuer H~P and Piilo J 2012 \href{http://dx.doi.org/10.1103/PhysRevA.85.052122}{{\em Phys. Rev. A\/} {\bf 85} 52122}.

\bibitem{werner}
Werner R~F 1989 \href{http://link.aps.org/doi/10.1103/PhysRevA.40.4277}{{\em Phys. Rev. A\/} {\bf 40} 4277}.

\bibitem{Mintert}
Mintert F, Carvalho A~R~R, Ku\'{s} M and Buchleitner A 2005 \href{http://www.sciencedirect.com/science/article/pii/S0370157305002334}{{\em Phys. Rep.\/}
  {\bf 415} 207}.

\bibitem{wootters}
Wootters W K 1998 \href{http://link.aps.org/doi/10.1103/PhysRevLett.80.2245}{{\em Phys. Rev. Lett.\/} {\bf 80} 2245}.

\bibitem{Modi}
Modi K, Brodutch A, Cable H, Paterek T and Vedral V 2012
\href{http://journals.aps.org/rmp/abstract/10.1103/RevModPhys.84.1655}{\textit{Rev. Mod. Phys.} \textbf{84} 1655}.

\bibitem{PhysRevLett.105.190502}
Daki\'{c} B, Vedral V and Brukner \v{C} 2010 \href{http://dx.doi.org/10.1103/PhysRevLett.105.190502}{{\em Phys. Rev. Lett.\/} {\bf 105} 190502}.

\bibitem{fano}
Fano U 1983 \href{http://dx.doi.org/10.1103/RevModPhys.55.855}{{\em Rev. Mod. Phys.\/} {\bf 55} 855}.

\bibitem{geometry}
Bengtsson I and Zyczkowski K 2008 {\em {Geometry of Quantum States}\/} (Cambridge University
  Press).
  
\bibitem{horodecki96}
Horodecki R and Horodecki M 1996 \href{http://dx.doi.org/10.1103/PhysRevA.54.1838}{{\em Phys. Rev. A\/} {\bf 54} 1838}.

\bibitem{meyer}
Meyer Jr C~D 1973 {\em SIAM J. Appl. Math.\/} {\bf 25} 597.  
  
\bibitem{Cubitt} Cubitt T S, Verstraete F, D\"{u}r W and Cirac J I 2003 \href{http://journals.aps.org/prl/abstract/10.1103/PhysRevLett.91.037902}{{\em Phys. Rev. Lett.\/} {\bf 91} 37902}.

\bibitem{PhysRevLett.108.250501}
Streltsov A, Kampermann H and Bru\ss{} D 2012 \href{http://link.aps.org/doi/10.1103/PhysRevLett.108.250501}{{\em Phys. Rev. Lett.\/} {\bf 108}
  250501}.

\bibitem{Chuan:2012kx}
Chuan T~K, Maillard J, Modi K, Paterek T, Paternostro M and Piani M 2012 \href{http://dx.doi.org/10.1103/PhysRevLett.109.070501}{{\em
  Phys. Rev. Lett.\/} {\bf 109} 70501}.

\bibitem{ex3} Fedrizzi A, Zuppardo M, Gillett G G, Broome M A, Almeida M P, Paternostro M, White A G and Paterek T 2013
 \href{http://link.aps.org/doi/10.1103/PhysRevLett.111.230504}{{\em Phys. Rev. Lett.\/} {\bf 111}
  230504}.

\bibitem{ex1} Vollmer C E, Schulze D, Eberle T, H\"{a}ndchen V, Fiur\'a\v{s}ek J and Schnabel R 2013
 \href{http://link.aps.org/doi/10.1103/PhysRevLett.111.230505}{{\em Phys. Rev. Lett.\/} {\bf 111}
  230505}.

\bibitem{ex2} Peuntinger C, Chille V, Mi\v{s}ta L Jr, Korolkova N, F\"{o}rtsch M, Korger J, Marquardt C and Leuchs G 2013
 \href{http://link.aps.org/doi/10.1103/PhysRevLett.111.230506}{{\em Phys. Rev. Lett.\/} {\bf 111}
  230506}.

\bibitem{Strel14}
Streltsov A, Kampermann H and Bru\ss{} D 2014 \href{http://journals.aps.org/pra/abstract/10.1103/PhysRevA.90.032323}{{\em Phys. Rev. A\/} {\bf 90}
  032323}.

\bibitem{PhysRevLett.106.160401}
Streltsov A, Kampermann H and Bru\ss{} D 2011 \href{http://link.aps.org/doi/10.1103/PhysRevLett.106.160401}{{\em Phys. Rev. Lett.\/} {\bf 106}
  160401}.

\bibitem{Piani2011} Piani M, Gharibian S, Adesso G, Calsamiglia J, Horodecki P and Winter A 2011 \href{http://journals.aps.org/prl/abstract/10.1103/PhysRevLett.106.220403}{\textit{Phys. Rev. Lett.} \textbf{106} 220403}.

\bibitem{Adesso} Adesso G, D'Ambrosio V, Nagali E, Piani M and Sciarrino F 2014 \href{http://journals.aps.org/prl/abstract/10.1103/PhysRevLett.112.140501}{\textit{Phys. Rev. Lett.} \textbf{112} 140501}.

\bibitem{Orieux} Orieux A, Ciampini M A, Mataloni P, Bru\ss{} D, Rossi M and Macchiavello C 2015 \href{http://journals.aps.org/prl/abstract/10.1103/PhysRevLett.115.160503}{\textit{Phys. Rev. Lett.} \textbf{115} 160503}.

\bibitem{Girolami} Girolami D, Souza A M, Giovannetti V, Tufarelli T, Filgueiras J G, Sarthour R S, Soares-Pinto D~O, Oliveira I S and Adesso G 2014 \href{http://journals.aps.org/prl/abstract/10.1103/PhysRevLett.112.210401}{\textit{Phys. Rev. Lett.} \textbf{112} 210401}.



\bibitem{Giovanetti} Giovannetti V, Lloyd S and Maccone L 2006 \href{http://journals.aps.org/prl/abstract/10.1103/PhysRevLett.96.010401}{\textit{Phys. Rev. Lett.} \textbf{96} 010401}.

\bibitem{Smerzi} Pezz\`{e} L and Smerzi A 2009 \href{http://journals.aps.org/prl/abstract/10.1103/PhysRevLett.102.100401}{\textit{Phys. Rev. Lett.} \textbf{102} 100401}.

\bibitem{Review} Pezz\`{e} L and Smerzi A 2014 in Tino G M and Kasevich M A (Eds.) \textit{Atom Interferometry. Proceedings of the International School of Physics Enrico Fermi, Course 188, Varenna}, 691--741, IOS Press.

\bibitem{shilov} Shilov G E 1977 {\em {Linear Algebra}\/} (Dover Publications).

\bibitem{0305-4470-34-47-329}
Verstraete F, Audenaert K, Dehaene J and {De Moor} B 2001 \href{http://iopscience.iop.org/0305-4470/34/47/329/}{{\em J. Phys. A\/}
  {\bf 34} 10327}.

\bibitem{Aravind19977}
Aravind P 1997 \href{http://www.sciencedirect.com/science/article/pii/S0375960197004386}{{\em Phys. Lett. A\/} {\bf 233} 7}.


\bibitem{robustwerner}
Li S-B and Xu J-B 2007 \href{http://link.springer.com/article/10.1140/epjd/e2006-00216-x}{{\em Euro. Phys. J. D\/} {\bf 41} 377}.

\bibitem{Palma}
Palma G M, Suominen K-A and Ekert A K 1996 \href{http://dx.doi.org/10.1098/rspa.1996.0029}{\textit{Proc. R. Soc. A} \textbf{452} 567}.

\bibitem{Duan}
Duan L-M and Guo G-C 1997 \href{http://journals.aps.org/prl/abstract/10.1103/PhysRevLett.79.1953}{\textit{Phys. Rev. Lett.} \textbf{79} 1953}.

\bibitem{Lidar}
Lidar D A, Chuang I L and Whaley K B 1998 \href{http://dx.doi.org/10.1103/PhysRevLett.81.2594}{\textit{Phys. Rev. Lett.} \textbf{81} 2594}.

\bibitem{Karpat}
Karpat G and Gedik Z 2011 \href{http://www.sciencedirect.com/science/article/pii/S0375960111012370}{\textit{Phys. Lett. A} \textbf{375}, 4166}.

\bibitem{eggeling}
Eggeling T and Werner R F 2001 \href{http://journals.aps.org/pra/abstract/10.1103/PhysRevA.63.042111}{\textit{Phys. Rev. A} \textbf{63} 042111}.

\bibitem{entdetection}
G\"{u}hne O and T\'{o}th G 2009 \href{http://www.sciencedirect.com/science/article/pii/S0370157309000623}{{\em Phys. Rep.\/} {\bf 474} 1}. 

\bibitem{TIENT}
Liu B-H \textit{et al.} \href{http://arxiv.org/abs/1603.09119}{arXiv:1603.09119v1}.

\end{thebibliography}
\end{document}